\documentclass[12pt]{iopart}

\usepackage{color}
\usepackage{psfrag}
\usepackage{graphicx}
\usepackage{epsfig}
\begin{document}

\title[Commensurability effects in Anderson localization]{Commensurability effects in one-dimensional Anderson localization:
anomalies in eigenfunction statistics}

\author{V.E.Kravtsov$^{1,2}$ and V.I.Yudson$^{3}$}

\address{$^1$The Abdus Salam International Centre for
Theoretical Physics, P.O.B. 586, 34100 Trieste, Italy.\\
$^{2}$Landau Institute for Theoretical Physics, 2 Kosygina
st.,117940 Moscow, Russia.\\$^{3}$Institute for Spectroscopy,
Russian Academy of Sciences, 142190 Troitsk, Moscow reg., Russia}
\begin{abstract}

 The one-dimensional (1d) Anderson model (AM), i.e. a tight-binding chain with
random uncorrelated on-site energies, has statistical anomalies at
any rational point $f=\frac{2a}{\lambda_{E}}$, where $a$ is the
lattice constant and $\lambda_{E}$ is the de Broglie wavelength. We
develop a regular approach to anomalous statistics of {\it
normalized eigenfunctions} $\psi(r)$ at such commensurability
points. The approach is based on an exact integral transfer-matrix
equation for a generating function $\Phi_{r}(u, \phi)$ ($u$ and
$\phi$ have a meaning of the squared amplitude and phase of
eigenfunctions, $r$ is the position of the observation point). This
generating function can be used to compute local statistics of
eigenfunctions of 1d AM at {\it any disorder} and to address the
problem of higher-order anomalies at $f=\frac{p}{q}$ with $q>2$. The
descender of the generating function ${\cal
P}_{r}(\phi)\equiv\Phi_{r}(u=0,\phi)$ is shown to be the
distribution function of phase which determines the Lyapunov
exponent and the local density of states.

In the leading order in the small disorder we have derived a
second-order partial differential equation for the $r$-independent
("zero-mode") component $\Phi(u, \phi)$ at the $E=0$
($f=\frac{1}{2}$) anomaly. This equation is nonseparable in
variables $u$ and $\phi$. Yet, we show that due to a hidden
symmetry, it is integrable and we construct an exact solution for
$\Phi(u, \phi)$ explicitly in quadratures. Using this solution we
have computed moments $I_{m}=N\langle |\psi|^{2m}\rangle$ ($m\geq
1$)  for a chain of the length $N \rightarrow \infty$ and found
an essential difference between their $m$-behavior in the
center-of-band anomaly and for energies outside this anomaly.
Outside the anomaly the "extrinsic" localization length defined
from the Lyapunov exponent coincides with that defined from the
inverse participation ratio ("intrinsic" localization length).
This is not the case at the $E=0$ anomaly where the extrinsic
localization length is smaller than the intrinsic one. At $E=0$ one
also observes an anomalous enhancement of large moments compatible
with existence of yet another, much smaller characteristic length scale.
\end{abstract}

\maketitle

\section{Introduction}\label{Intro}
Concepts and methods of the localization theory, which counts its
origin from the seminal Anderson paper \cite{Anderson}, have
penetrated almost all branches of modern
physics\cite{Mirlin2008,Mirlin2000} - from the description of
transport in disordered media and the Quantum Hall effect to the
theory of chaotic systems \cite{Modugno} and turbulence.
The one-dimensional (1d) Anderson model \cite{Anderson} (AM) -- a
tight-binding model with a diagonal disorder -- is determined by the
Schr\"{o}dinger equation for the particle wave function $\psi_{i}$
at a site $i$
\begin{equation}
\label{SE}
t[\psi_{i-1} +\psi_{i+1}] + \varepsilon_i\psi_{i} = E\psi_{i} \, .
\end{equation}
Here, the nearest neighbor hopping amplitude $t$ is the same for all
bonds (below we put $t=1$); the on-site energy $\varepsilon_{i}$ is
a random variable uncorrelated at different sites and characterized
by a zero average $\langle \varepsilon_{i} \rangle=0$ and the
variance
\begin{equation}\label{variance}
\langle \varepsilon_{i}\varepsilon_{j}\rangle=\sigma^{2}\delta_{ij}
\, ,
\end{equation}
which is a measure of the disorder strength. The variance
Eq.(\ref{variance}) is the only quantity that enter our theory at
weak disorder $\sigma^{2}\ll 1$. For a generic on-site disorder we
derive equations in terms of the entire distribution function of the
on-site energies (uncorrelated at different sites):
\begin{equation}
\label{ent-dist} {\cal F}(\varepsilon)=\langle
\delta(\epsilon-\epsilon_{i})\rangle.
\end{equation}
For a finite chain ($i = 1, 2, \cdots, N$), Eq.(\ref{SE}) is
supplemented with the definition
\begin{equation}
\label{BC} \psi_0 = 0 = \psi_{N+1},
\end{equation}
which is equivalent to the hard-wall boundary conditions. In the
absence of the disorder ($\varepsilon \equiv 0$), the normalized
wave functions of the chain are $\psi_j = \sqrt{2/N}\sin{(kj)}$ and
the corresponding eigenenergies are
\begin{equation}
\label{E-k}
E(k) = 2\cos{(k)} \, ,
\end{equation}
where $k = \pi l/N$ ($l=1,..., N$); for an infinite chain ($N \rightarrow \infty$) it fills the
interval $(0, \pi)$. Due to the band symmetry, it is sufficient to consider only $k \in (0, \pi/2)$.

The most studied is the continuous limit of AM, where the lattice
constant $a\rightarrow 0$ at $ta^{2}$ remaining finite
\cite{Borland,Halperin,Ber,AR,Mel,Kolok,Pendry-rev}. There was also a great deal of activity
\cite{1dRev, Pastur} aimed at a rigorous mathematical description of
1d AM. However, despite considerable efforts invested, a lot of
issues concerning 1d AM still remain unsolved. Among them there are
effects of commensurability between the de-Broglie wavelength
$\lambda_{E}$ (dependent on the energy $E$) and the lattice constant
$a$, i.e. the anomalous behavior at rational values of $f=k/\pi$.

The first known manifestation of commensurability effects was found
and described quite early \cite{Wegner,Derrida} for the simplest
objects - the density of states (DoS) and the Lyapunov exponent.
The latter is defined in terms of a solution of the Schr\"{o}dinger
equation (\ref{SE}) for a
\emph{semi-infinite} chain ($i = 1, 2, \cdots$) supplemented with
the definition $\psi_0 = 0$ at only one end. For an arbitrary energy
$E$ and a generic boundary condition $\psi_1 \sim 1$, the solution
to Eq.(\ref{SE}) is a superposition of two solutions, decreasing and
increasing with the increase of $i$. The increasing part determines
the Lyapunov exponent $\gamma(E)$ and the corresponding localization
length $\ell(E)$:
\begin{equation}\label{Lyapunov}
\hspace{-1.5cm} \frac{1}{\ell(E)} = \Re \, \gamma(E) =
\lim_{N\rightarrow \infty}\frac{1}{N} \Re \, \log{(\psi_N/\psi_{1})}
= \lim_{N\rightarrow \infty}\frac{1}{N}\Re \,
\sum^{N}_{j=2}\log{(\psi_j/\psi_{j-1})} \, .
\end{equation}
In the continuous model with a weak disorder $\sigma^{2}\ll 1$, the
Lyapunov exponent $\gamma^{0}(E)$ and the localization length
\begin{equation}
\label{ell0} \ell_{0}(E)=\frac{8\sin^2{k}}{\sigma^{2}},
\end{equation}
 are smooth
functions of energy $E$. However, according to
\cite{Wegner,Derrida}, for a discrete chain, the functions
$\gamma(E)$ and $\ell(E)$ possess anomalous deviations from
$\gamma^{0}(E)$ and $\ell_{0}(E)$ in narrow windows of the size
$\propto w$  around the points $k=\pi/2$ (i.e., $E=E(k)=0$) and
$k=\pi/3$ (i.e., E= $E(k)=1$). The Lyapunov exponent sharply {\it
decreases} at $k=\pi/2$ (which is usually associated with an {\it
increase} of the localization length) but may both increase or
decrease at $k=\pi/3$ depending on the third moment $\langle
\varepsilon^3_{i}\rangle$ of the on-site energy distribution
\cite{Derrida}. It was also conjectured \cite{Derrida} that
progressively weakening anomalies (for the weak disorder) may take
place at every rational point $k/\pi = m/n$ (with natural $m$ and
$n$) of the band.

More recently \cite{Titov, AL} it has been found that also the
statistics of conductance in 1d AM are anomalous at the center of
band ($E=0$, $k=\pi/2$). This is not too surprising, because the
chain conductance is expressed via the reflection and transition
coefficients of an electron wave coming from a perfect leads to
\emph{one} of the ends of the disordered chain. This problem has a
lot of similarities with the problem of calculation of the Lyapunov
exponent: solutions to both of them are expressed in terms of a
probability distribution function $\mathcal{P}(\phi)$ of  the phase
$\phi$. The latter can be interpreted as a ``phase" parameter in the
representation of the wave function $\psi_j $ in the form $\psi_j =
a_j\cos(kj + \phi_j)$ with slowly varying real amplitude and phase.
The center of band anomaly corresponds to an emergent non-triviality
of the distribution function $\mathcal{P}(\phi)$ as compared to the
trivial isotropic angular distribution in the continuous problem.

In a sense, the both problems touch ``extrinsic" properties of
localization, as the Lyapunov exponent describes only the {\it
tails} of localized wave functions.  In the problem of the average
logarithm of  conductance, the extrinsic character of this quantity
is set by the distance $L\gg \ell(E)$ between the ideal leads. In
other applications of which most important is the interplay between
the localization and non-linearity \cite{Modugno}, one is interested
in the number of sites with a high amplitude of the wave function.
This "intrinsic" picture of localization is better represented by
the   {\it inverse participation ratio} (IPR) $I \sim \int
dx\;\langle|\psi(x)|^{4}\rangle$, where $\psi(x)$ is a random {\it
normalized} eigenfunction obeying the Shr\"{o}dinger equation
Eq.(\ref{SE}) with the boundary conditions Eq.(\ref{BC}) at the
\emph{both ends} of the chain. The relation between the two
descriptions is not clear yet. In particular, it is not known
whether the localization length, determined via IPR and sensitive to
short-range characteristics of eigenfunctions, coincides at points
of anomaly $k/\pi = m/n$ with the length determined via Lyapunov
exponent Eq.(\ref{Lyapunov}). This question is one of the
motivations of the present paper aimed to develop a formalism to
tackle a class of ``intrinsic" problems connected with statistics of
\emph{normalized} eigenfunctions.

Here it is worth noting that IPR and other quantities determined by
eigenfuctions, are much more difficult to compute than the
Lyapunov exponent. The reason is that the latter problem does not
require a {\it stationary} solution to the Shr\"{o}dinger equation
neither to obey the boundary conditions at the both ends, nor to be
normalized. An attack on the ``two-end"
 problem was undertaken in the pioneering paper \cite{Wegner} but with only limited success. The
 normalization condition imposed on the wave function amplitudes $\{a_j\}$ turned out to be very
 difficult for an analytical treatment. Only few quantities which are effectively insensitive to
 the normalization constraint have been calculated: they are the averaged density of states (DOS)
  and the ratios of wave functions at different sites.

The difficulty of treating IPR and other quantities determined by
local eigenfunctions is due to the fact that one needs to deal with
an unknown joint probability distribution function $P(u,\phi)$ of
two variables, the phase and the amplitude.

In this paper we develop a regular approach to treat the anomalies
in the eigenfunction statistics. The approach is based on a
transfer-matrix equation (TME) for a generating function of two
variables $\Phi(u, \phi)$ ($u$ has a meaning of the squared
amplitude of wave functions), which is a universal tool to describe
properties of a generic 1d or quasi-1d system. The generating
function can be used to compute {\it any} local statistics of {\it
normalized} eigenfunctions of 1d AM. It also determines a {\it joint
probability distribution} $P(u,\phi)$ of the (squared)
amplitude $u$ and phase $\phi$ of the random eigenfunction $\psi\sim
\sqrt{u}\,\cos{\phi}$.

We will concentrate mostly on the study of the principle
(``center-of-band") anomaly at $k=\pi/2$ ($E(k) =0$) at weak
disorder and show that the corresponding TME for $\Phi(u, \phi)$ has
anomalous terms which make it essentially two-dimensional
second-order partial differential equation (PDE). This equation is
nonseparable in variables $u$ and phase $\phi$. Yet, we show that it
is integrable and we construct its exact solution explicitly in
quadratures.

In the next section \ref{TME-derivation} we present an
\emph{elementary } derivation of the transfer-matrix equation (TME)
 which is valid at any strength of disorder.  This derivation does
not exploit the supersymmetry method \cite{Efet-book}, used in
earlier approaches (see \cite{Mirlin-Fyodorov-1991,Mirlin2000,OsK}).
By the same token we derive the general expression for the
statistical moments of $|\psi|^{2}$ and the mean local density of
states in terms of the generating function.  In section
\ref{Ampl-phase-variables} we introduce amplitude-phase variables
connected with the representation of eigenfunctions in the form
$\psi \sim \sqrt{u}\cos{\phi}$. In these variables, assuming weak
disorder $\sigma^{2}\ll 1$, we obtain a partial differential
equation for the generating function, section \ref{PDE for GF}. From
that moment on we concentrate on the study of the principle,
center-of-band anomaly. In section \ref{Center-of-band-anomaly} we
consider a partial differential equation for the case $k=\pi/2$
($E(k) =0$) and show its integrability. Namely, we show that this
equation can be factorized in new variables and we can construct its
solutions. But it turns out that due to non-Hermitian nature of the
differential operator, there is a continuum of possible solutions.
This huge redundancy problem is analyzed in section \ref{Solution
for GF} where we show how physical requirements imposed on the
generating function allow to find the unique solution. This solution
is used in section \ref{Physical applications} to compute
statistical moments of $|\psi|^{2}$. It is shown that these moments
cannot be described by only one-parameter. This invalidate  the
one-parameter scaling at the anomalous center-of-band point $E=0$.
In Conclusion we  point out to the analogies with certain dynamical
systems and discuss the open problems.

\section{Elementary derivation of the transfer-matrix equation (TME)}\label{TME-derivation}

The quantity
\begin{equation}\label{IPR-m-definition}
I_{m}(r,E) \equiv \frac{1}{\nu(E)}\sum_{\nu}
\left\langle|\psi^{(\nu)}_{r}|^{2m} \delta(E-E_{\nu})\right\rangle
\,
\end{equation}
generalizes the concept of the inverse participation ratio (IPR)(
given by Eq.(\ref{IPR-m-definition}) at $m=2$) to an arbitrary
natural $m$. Here, $\psi^{(\nu)}_{r}$ is a $\nu$-th eigenfunction of
Eqs.(\ref{SE}) and (\ref{BC}) with the eigenenergy $E_{\nu}$, the
summation runs over all states, the angular brackets denote the
averaging over the ensemble of random site potentials
$\{\varepsilon_j\}$, and the quantity
\begin{equation}\label{DOS-definition}
 \nu(E)=N^{-1}\sum_{\nu} \langle \delta(E-E_{\nu})\rangle \,
\end{equation}
 is the averaged density of states (DoS) at a given energy $E$.
 For a weak disorder $\sigma^{2}\ll 1$, the DoS $\nu(E)$ inside the energy band and outside of ``anomalous regions"
 only slightly (by $O(\sigma^{2})$) differs from the corresponding function $\nu_0(E)$ for the ordered system:
\begin{equation}\label{DOS}
\nu(E) \approx \nu_0(E)= \frac{1}{2\pi \sin{[k(E)]}} \, .
\end{equation}
However, the difference is known \cite{Wegner,Derrida} to be
appreciable ($\sim O(1)$) near the $E=0$ anomaly (see section
\ref{DOS-section}).

For $m=1$, the quantity $\sum_{r}I_{m=1}(r,E) = N$ just due to the
normalization of wave functions. This implies that
$I_{m=1}(E)=1+O(1/N)$, as far from the ends of a long chain
$I_{m}(r,E)$ is independent of the position $r$.

For $m \geq 2$, the quantity $I_{m}(r,E)$ can be expressed in terms
of the Green's functions of the problem, Eqs.(\ref{SE}) and
(\ref{BC}), with the use of a limiting procedure (see, e.g., the
review \cite{Mirlin2000}):
\begin{equation}\label{IPR-m-representation}
I_{m}(r,E) = \lim_{\eta\rightarrow +0} \frac{(i\eta)^{m-1}}{2\pi
\nu(E)} \langle G_{r,r}^{m-1}(E_{+}) G_{r,r}(E_{-})\rangle \,\,\, ;
\,\,\, E_{\pm} = E \pm \frac{i\eta}{2} \, .
\end{equation}
The Green's functions
$$G_{j,r}(E_{\pm}) = \sum_{\nu} \frac{\psi^{(\nu)}_{j}\psi^{(\nu)*}_{r}} {E_{\pm}-E_{\nu}}$$
(with a source at the site $r$) obey the equations
\begin{eqnarray}
[G_{j-1,r} + G_{j+1,r}] + \varepsilon_j G_{j,r} + \delta_{j,r} = E_{\pm} G_{j,r} \, \label{SE-Green}\\
G_{0,r} = 0 = G_{N+1,r} \, . \label{BC-Green}
\end{eqnarray}
Instead of the supersymmetry approach \cite{Efet-book}, where
Green's functions are represented as functional integrals over the
usual complex (``bosonic") and Grassmann (``fermionic") variables,
here we will present an elementary derivation. The derivation is
close in spirit to the methods used in refs. \cite{Wegner,Derrida}.

For $j \neq r$, dividing Eq.(\ref{SE-Green}) by $G_{j,r}$ (as is justified below, this quantity
differs from 0), we obtain the recursive equation:
\begin{eqnarray}\label{SE-q}
q^{\pm}_{j} + \frac{1}{q^{\pm}_{j-1}} \equiv \frac{G_{j+1,r}(E_{\pm})}{G_{j,r}(E_{\pm})} +
\frac{G_{j-1,r}(E_{\pm})}{G_{j,r}(E_{\pm})} = E_{\pm} - \varepsilon_j \, ,
\end{eqnarray}
supplemented with the definitions (see Eq.(\ref{BC-Green}))
\begin{eqnarray}\label{BC-q}
1/q^{\pm}_{0} = 0 = q^{\pm}_{N} \, ,
\end{eqnarray}
Starting with $j=1$ and using Eq.(\ref{SE-q}) to go from $j-1$ to $j$, one can find all the
$q^{\pm}_j$, $0<j<r$ as functions of $\varepsilon_1, \ldots, \varepsilon_{r-1}$. Similarly,
starting with $j=N$ and going from $j$ to $j-1$, one finds all the $q^{\pm}_j$, $r \leq j<N$
as functions of $\varepsilon_{r+1}, \ldots, \varepsilon_{N}$. Finally, from Eq.(\ref{SE-Green})
 at $j=r$ we obtain the quantities of our interest, $G_{r,r}(E_{\pm})$:
\begin{eqnarray}\label{G-q}
\hspace{-2.5cm} G_{r,r}(E_{\pm}) = \frac{1}{E_{\pm} - \varepsilon_r - q^{\pm}_{r} - 1/q^{\pm}_{r-1}}
= \mp i\int^{\infty}_{0}{ d\lambda \, \exp{\left[\pm i\lambda\left(E_{\pm} - \varepsilon_r -
q^{\pm}_{r} - \frac{1}{q^{\pm}_{r-1}}\right)\right]}}  ;\\
\hspace{-2.5cm} G^{m-1}_{r,r}(E_{+}) = \frac{(-i)^{m-1}}{(m-2)!}\int^{\infty}_{0}{ d\lambda \,
\lambda^{m-2}\exp{\left[i\lambda\left(E_{+} - \varepsilon_r - q^{+}_{r} - \frac{1}{q^{+}_{r-1}}\right)\right]}}
\,\,\, ; \,\,\, m \geq 2
\, .
\end{eqnarray}
To justify the transition from Eq.(\ref{SE-Green}) to Eq.(\ref{SE-q}), we should
prove that $G_{j,r}(E_{\pm}) \neq 0$. Note that $q^{+}_1 = E + i\eta/2 - \varepsilon_1$,
hence $\Im \, q^{+}_1 > 0$. Assuming that $\Im \, q^{+}_{j-1} > 0$, we obtain (for $j <r$):
$\Im\, q^{+}_{j} = - \Im\, (1/q^{+}_{j-1}) + \eta/2 > 0$, hence $\Im \, q^{+}_{j} > 0$,
and by induction this is true for any $j < r$, which means all the corresponding $q^{+}_j$
(and $G_{j,r}$) are nonzero. Similarly, for quantities $q^{-}_j$ we find $\Im \, q^{-}_{j} < 0$.
And for the case $j > r$, we obtain $\Im \, q^{+}_j < 0$ and $\Im \, q^{-}_j > 0$.

As a by-product of this proof, we have confirmed the expected positiveness (negativeness) of the
imaginary part of the denominator of $G_{r,r}(E_{+})$ ($G_{r,r}(E_{-})$), which justifies the integral
representation in Eq.(\ref{G-q}).

Using Eq.(\ref{G-q}), the expression Eq.(\ref{IPR-m-representation})
can be represented in the form ($ m \geq 2$):
\begin{eqnarray}\label{IPR-m-integrals}
\hspace{-2.cm} I_{m}(r,E) = \frac{1}{2\pi(m-2)! \, \nu(E)}
&&\lim_{\eta\rightarrow +0} \eta^{m-1} \int^{\infty}_{0}
\int^{\infty}_{0} d\lambda_1 d\lambda_2  \lambda^{m-2}_1 \,\,
\mathrm{e}^{i(\lambda_1 -\lambda_2)E - (\lambda_1 + \lambda_2)\eta/2} \nonumber \\
&&\langle \mathrm{e}^{-i(\lambda_1 -\lambda_2)\varepsilon_r} \rangle
 \mathcal{R}_{r-1}(\lambda_1,\lambda_2)\tilde{\mathcal{R}}_{r}(\lambda_1,\lambda_2)
\, .
\end{eqnarray}
Here
\begin{eqnarray}
&&\mathcal{R}_{j}(\lambda_1,\lambda_2) \equiv \langle \exp{[-i\lambda_1/q^{+}_j + i\lambda_2/ q^{-}_j]}
\rangle  \,\,\, ; \,\,\,\,\,\,\,\,  j < r \, ; \label{R-definition} \\
&&\tilde{\mathcal{R}}_{j}(\lambda_1,\lambda_2) \equiv \langle \exp{[-i\lambda_1 q^{+}_j +  i\lambda_2 q^{-}_j]}
\rangle  \,\,\, ; \,\,\,  r \leq j \, , \label{R-tilde-definition}
\end{eqnarray}
where the averaging in Eqs.(\ref{R-definition}) and
(\ref{R-tilde-definition}) is performed over random energies
$\varepsilon_1, \ldots, \varepsilon_{r-1}$ and $\varepsilon_{r+1},
\ldots, \varepsilon_{N}$, respectively.  The crucial assumption here
is that the random energies $\varepsilon_{r}$ are uncorrelated at
different sites. The functions $\mathcal{R}_{j}$ and
$\tilde{\mathcal{R}}_{j}$ obey recurrent equations. To derive them,
we use the following identity for the Bessel function $J_0(x)$:
\begin{eqnarray}\label{identity}
\exp{(-z)} = - \int^{\infty}_{0} d\lambda' J_0(2\sqrt{\lambda'})
\frac{\partial}{\partial \lambda'}\exp{(-\lambda'/z)} \,\,\,\,\,\,\,\,\,\,\ (\Re \, z > 0 ) \,\,\,\, ,
\end{eqnarray}
which allows us to convert the factor $1/q_j$ to $q_j$ in the
exponent of Eq.(\ref{R-definition})   and then to apply
Eq.(\ref{SE-q}). As a result, we obtain:
\begin{eqnarray}\label{R-equation}
\hspace{-1.0cm}
\mathcal{R}_{j}(\lambda_1,\lambda_2) = \int^{\infty}_{0} \int^{\infty}_{0} &&d\lambda'_1 d\lambda'_2
\, J_0(2\sqrt{\lambda_1\lambda'_1})J_0(2\sqrt{\lambda_2\lambda'_2}) \nonumber \\
&&\frac{\partial}{\partial \lambda'_1}\frac{\partial}{\partial \lambda'_2}\left[
\left\langle \mathrm{e}^{i(\lambda'_1 -\lambda'_2)(E-\varepsilon_j)}\right\rangle
\mathrm{e}^{- (\lambda'_1 + \lambda'_2)\eta/2}
\mathcal{R}_{j-1}(\lambda'_1,\lambda'_2) \right]
\,
\end{eqnarray}
with the initial condition $\mathcal{R}_{0}(\lambda_1,\lambda_2) = 1$.
Note that the function $\mathrm{e}^{-i(\lambda'_1 -\lambda'_2)\varepsilon_{j}}$ in the integrand is statistically independent of quantities
$\varepsilon_{j-1}, \varepsilon_{j-2}$, etc. which determine the function $\mathcal{R}_{j-1}(\lambda'_1,\lambda'_2)$.
In a similar way but proceeding from the site $N$ to the site $j$, we derive the recursive equation for
$\tilde{\mathcal{R}}_{j}(\lambda_1,\lambda_2)$:
\begin{eqnarray}\label{R-tilda-equation}
\hspace{-1.0cm}
\tilde{\mathcal{R}_{j}}(\lambda_1,\lambda_2) = \int^{\infty}_{0} \int^{\infty}_{0} &&d\lambda'_1 d\lambda'_2 \,
J_0(2\sqrt{\lambda_1\lambda'_1})J_0(2\sqrt{\lambda_2\lambda'_2}) \nonumber \\
&&\frac{\partial}{\partial \lambda'_1}\frac{\partial}{\partial \lambda'_2}\left[
\left\langle \mathrm{e}^{i(\lambda'_1 -\lambda'_2)(E-\varepsilon_{j+1})}\right\rangle
\mathrm{e}^{- (\lambda'_1 + \lambda'_2)\eta/2}
\tilde{\mathcal{R}}_{j+1}(\lambda'_1,\lambda'_2) \right]
\,
\end{eqnarray}
with the initial condition $\tilde{\mathcal{R}}_{N}(\lambda_1,\lambda_2) = 1$.
For the considered case of site-independent statistics of the local disorder,
one can see immediately that the function $\mathcal{R}_{N-j}(\lambda_1,\lambda_2)$ obeys the
recursive Eq.(\ref{R-tilda-equation}) and equals unity at $j=N$. Therefore, this function
should coincide with the function $\tilde{\mathcal{R}}_{j}(\lambda_1,\lambda_2)$ and we arrive
at an identity:
\begin{eqnarray}\label{R-tilda-R-identity}
\tilde{\mathcal{R}_{j}}(\lambda_1,\lambda_2) = \mathcal{R}_{N-j}(\lambda_1,\lambda_2)
\, .
\end{eqnarray}
To perform the limit operation $\eta\rightarrow +0$ in
Eq.(\ref{IPR-m-integrals}) and Eq.(\ref{R-equation}), we introduce
the new variables
\begin{eqnarray}\label{s-and-v}
 s=\eta (\lambda_1 + \lambda_2)/2  \,\,\, ;  \,\,\, v = \lambda_1 - \lambda_2 \,
\end{eqnarray}
and the new functions:
\begin{eqnarray}\label{W-definition}
\hspace{-2.cm}
W_{j}(s,v) \equiv \mathcal{R}_{j}(s/\eta + v/2, s/\eta - v/2) \,\, ; \,\,
\tilde{\mathcal{R}}_{j}(s/\eta + v/2, s/\eta - v/2) = W_{N-j}(s,v)\, .
\end{eqnarray}
Using asymptotic form of the Bessel function, integrating by parts
in Eq.(\ref{R-equation}), and neglecting infinitely fast oscillating
terms, we arrive at the following recursive equation:
\begin{eqnarray}\label{W-equation}
\hspace{-2.5cm}
W_{j}(s,v) = \frac{\sqrt{s}}{2\pi}\int^{\infty}_{-\infty}dv'\int^{\infty}_{0} \frac{ds'}{(s')^{3/2}} \mathrm{e}^{-s'}
\cos{\left[\sqrt{ss'}\left(\frac{v}{s} + \frac{v'}{s'} \right)\right]}\mathrm{e}^{iv'E}
 \,\chi(v')\, W_{j-1}(s',v'),
\end{eqnarray}
where $\chi(v')$ is the characteristic function of the on-site
energy distribution $\mathcal{F}(\varepsilon)$ Eq.(\ref{ent-dist}):
\begin{equation}
\label{char} \chi(v)=\int d\varepsilon\,{\cal
F}(\varepsilon)\,e^{-i\,\varepsilon\,v}\equiv\left\langle
\mathrm{e}^{-iv'\varepsilon_j}\right\rangle.
\end{equation}

 In the new variables, Eq.(\ref{IPR-m-integrals})
takes the form ($m \geq 2$):
\begin{eqnarray}\label{IPR-m-integrals-new-s-v}
\hspace{-2.5cm} I_{m}(r,E) = \frac{1}{2\pi(m-2)!\, \nu(E)}
\int^{\infty}_{-\infty} dv \int^{\infty}_{0} d s \,
s^{m-2}\,\mathrm{e}^{ivE -s}  \,\chi(v)\, W_{r-1}(s,v)W_{N-r}(s,v)
\, .
\end{eqnarray}
Free of any limit operation, Eqs.(\ref{W-equation}) and (\ref{IPR-m-integrals-new-s-v})
are the starting point of our analysis. For the 1d problem of interest, they were obtained in \cite{OsK},
and still earlier in \cite{Mirlin-Fyodorov-1991,A-C-Anderson-Thouless-1973} in a study
of the localization transition on the Bethe lattice.
The presented here elementary derivation of these equations is considerably simpler than
the supersymmetry approach used in \cite{OsK,Mirlin-Fyodorov-1991}. Also, it allows
to establish a relation between the generating function $W_j(s,v)$ of our interest
and the phase distribution function $\mathcal{P}(\phi)$ (see below).

\section{The ``amplitude-phase" variables $z$ and $\phi$}\label{Ampl-phase-variables}
\subsection{Exact equations in $(s,q)$ and ($z,\phi$) variables}
To proceed, we introduce  the Fourier-transform of $W_j(s,v)$ in the
variable $v$:
\begin{eqnarray}\label{W-Fourier-definition}
\tilde{W}_j(s,q) = \int dv \, \mathrm{e}^{iqv} \, W_j(s,v) \, .
\end{eqnarray}
The basic equations (\ref{IPR-m-integrals-new-s-v}),(\ref{W-equation}) in
the new variables $(s,q)$ take the following form:
 \begin{eqnarray}\label{IPR-m-integrals-new}
&&I_{m}(r,E) = \frac{1}{2\pi(m-2)!\, \nu(E)}
\int^{\infty}_{0} d s \, s^{m-2}\,\mathrm{e}^{-s} \nonumber \\
&&\int^{\infty}_{-\infty} \frac{d q \, d q'}{2\pi}\,
\tilde{W}_{r-1}(s,q)\,{\cal F}(E-q-q')\,\tilde{W}_{N-r}(s, q') \, ,
\end{eqnarray}
\begin{eqnarray}\label{W-equation-Fourie}
\tilde{W}_{j}(s,q) = \frac{\mathrm{e}^{-sq^2}}{q^2}
\int^{\infty}_{-\infty} d q' \, {\cal
F}(E-q^{-1}-q')\,\tilde{W}_{j-1}(sq^2, q') \, .
\end{eqnarray}
Eqs.(\ref{IPR-m-integrals-new}),(\ref{W-equation-Fourie}) are {\it
exact} for uncorrelated on-site energies with the arbitrary
distribution function ${\cal F}(\varepsilon)$.

Now we introduce yet another couple of variables, $z$ and $\phi$, determined by:
\begin{eqnarray}\label{z-q-variables}
s = z\cos^2{(\phi +k)} \,\,\,\,\,\,\ ; \,\,\, \,\,\,\, q = \frac{\cos{\phi}}{\cos{(\phi + k)}}
\, ,
\end{eqnarray}
and a new function:
\begin{eqnarray}\label{Phi-definition}
\Phi_j(z,\phi) = \tilde{W}_j(s,q)\frac{\sin{k}}{2\pi\cos^2{(\phi +
k)}},
\end{eqnarray}
with the boundary condition at $j=0$
\begin{eqnarray}\label{BC-z-phi}
W_{0}(s,v) = &&\mathcal{R}_{0}(s/\eta + v/2, s/\eta - v/2 ) = 1 \,\,  \Rightarrow  \nonumber \\
&& \Rightarrow  \, \Phi_{0}(z,\phi) =
2\pi\delta(q)\frac{\sin{k}}{2\pi \cos^2{(\phi+k)}} = \delta(\phi -
\pi/2) \,  .
\end{eqnarray}
Here $z \in (0, \infty)$; $k$ is determined by the relation $E =
2\cos{k}$; and the ``phase" variable $\phi$ changes within the
interval $(0, \pi)$, where there is one-to-one correspondence
between $\phi$ and $q(\phi)$. Alternatively, we may use so called
``extended band" representation, where $\phi$ is arbitrary, but the
function $\Phi_j(z,\phi)$ obeys the periodicity condition
\begin{eqnarray}\label{Phi-periodicity}
\Phi_j(z,\phi +\pi) = \Phi_j(z,\phi ) \,\,\,\, ; \,\,\,  \forall \phi
\, ,
\end{eqnarray}
and integrations over $\phi$ can be taken over any interval of the
length $\pi$.

In new variables, exact Eq.(\ref{IPR-m-integrals-new}) for moments reads
\begin{eqnarray}\label{IPR-m-integrals-z-phi-exact}
&&\hspace{-2.cm}I_{m}(r,E) = \frac{1}{(m-2)!\nu(E)} \int^{\pi}_{0}
d\phi \, d \phi' \cos^{2(m-1)}({\phi}) \int^{\infty}_{0} d z \, z^{m-2}\mathrm{e}^{-z\cos^{2}{(\phi+k)}}
\nonumber \\
&&\hspace{-2.cm}{\cal F}\left(E-\frac{\cos{\phi}}{\cos{(\phi + k)}} - \frac{\cos{\phi'}}{\cos{(\phi' + k)}}\right)
\Phi_{r-1}(z,\phi)\, \Phi_{N-r}\left(z\frac{\cos^{2}{(\phi+k)}}{\cos^{2}{(\phi'+k)}}\,, \phi'\right) \, ,
\end{eqnarray}
while Eq.(\ref{W-equation-Fourie}) takes the form:
\begin{eqnarray}\label{Phi-equation}
\hspace{-2.5cm} \Phi_{j+1}(z,\phi) =
\frac{\sin{k}\,\mathrm{e}^{-z\cos^2{\phi}}}{ \cos^2{\phi}}
\int^{\pi}_{0} d\phi' {\cal F}\left((\sin k\,(\tan{\phi'} -
\tan{\phi})\right)\,
\Phi_{j}\left(\frac{z\cos^{2}\phi}{\cos^{2}\phi'},\phi'-k \right) \, .
\end{eqnarray}
From now on the function $\Phi_{j}(z,\phi)$ will be referred to as
the \emph{generating function} which determines the statistical
moments of $|\psi|^{2}$ distribution.

\subsection{The Lyapunov and the reflection phase.}
The integrand in Eq.(\ref{IPR-m-integrals-z-phi-exact}) for the
$m$-th moment of the quantum-mechanical probability density
$I_{m}\sim\langle |\psi^{2}|^{m}\rangle $ contains
$\cos^{2m}(\phi)\,z^{m}$ which suggests the physical meaning of
$\phi$ and $\sqrt{z}$ as a phase and the amplitude of the
wave function $\psi\propto \sqrt{z}\,\cos(\phi)$. However, one may
ask a question what is the meaning of a phase for a wave function in
one dimensions which may always be chosen {\it real}. We answer this
question below and show that the distribution of phase $\phi$ is related with
that of the phase of coefficient $r=|r|e^{i\theta}$ of reflection from
a semi-infinite disordered chain \cite{Barnes&Luck-1990}.

To this aim, we follow Kappus and Wegner
\cite{Wegner} and introduce $a_j > 0$ and $\phi_j$ variables defined
on the link between the sites $j$ ($0< j < r$) and $j+1$ in such a
way that (for a fixed site $r$)
\begin{eqnarray}\label{links}
\hspace{-1.5cm}
G_{j,r}(E) = a_{j}\cos{\phi_j} \,\,\, ; \,\,\,
G_{j+1,r}(E) = a_{j}\cos{(\phi_j +k)} \,\,\,
\Rightarrow \,\,\,
q_{j} = \frac{\cos{(\phi_j +k)}}{\cos{\phi_j}} \, .
\end{eqnarray}
One can see that such an ansatz is compatible with Eqs.(\ref{SE-Green})-(\ref{SE-q}).
For the link between the sites 0 and 1 we define $\phi_0 = \pi/2$,
so that $1/q_0 \equiv 0$. For brevity, superscripts $\pm$ in
Eq.(\ref{links}) are not indicated.

Now we use the well-known expression for the Green's function in
terms of the solutions $\psi^{<}$ and $\psi^{>}$ to the Schroedinger
equation Eq.(\ref{SE}) with the {\it arbitrary} energy $E$ which
obey only {\it one} of the two boundary conditions: the function
$\psi^{<}$ obeying the boundary condition $\psi^{<}(0)=0$, while the
function $\psi^{>}$ obeying $\psi^{>}(N+1)=0$:
\begin{equation}
\label{gele}
G_{r,r'}(E)=W_{E}^{-1}\,\left\{\matrix{\psi^{<}_{E}(r)\,\psi^{>}_{E}(r'),
& r<r' \cr \psi^{>}_{E}(r)\,\psi^{<}_{E}(r'), & r>r'}\right.,
\end{equation}
where
$W_{E}=\psi^{<}_{E}(r)\,\psi^{>}_{E}(r+1)-\psi^{>}_{E}(r)\,\psi^{<}_{E}(r+1)$
is the Wronskian.

Then Eq.(\ref{links}) is equivalent to
\begin{eqnarray}
\label{phi-def} \hspace{-1.0cm} \psi^{<}_{E}(j)=
\sqrt{z_{j}}\cos{\phi_j} \,\,\, ; \,\,\, \psi^{<}_{E}(j+1)=
\sqrt{z_{j}}\cos{(\phi_j +k)} .
\end{eqnarray}
Eq.(\ref{phi-def}) gives the definition of the variables $z$ and
$\phi$ in terms of the solution to the Schroedinger Eq.(\ref{SE})
with the boundary condition at the left end. As was explained above
this is exactly the formulation of the problem of Lyapunov exponent.
That is why we will refer to the phase $\phi$ as the "Lyapunov
phase". In the absence of disorder the relationship
Eq.(\ref{phi-def}) is natural, as the shift of phase between the
$j$-th and $j+1$-th site is indeed equal to $k$. We will see later
on that the so defined phase has a flat distribution ${\cal
P}(\phi)=\frac{1}{\pi}$ far from the end ($j\gg \ell(E)$) of a
weakly disordered chain unless the energy $E$ is close to the
anomalous points $k(E)=\pi\,p/q$.

The probability distribution $\mathcal{P}(\phi)$ of the Lyapunov phase
is related with the distribution $P_{\mathrm{ref}}(\theta)$ (derived
in \cite{Barnes&Luck-1990}) of the phase of the coefficient of reflection $r_{N}=|r_{N}|\mathrm{e}^{i\theta_{N}}$ from a long chain
(of size $N \rightarrow\infty$).
Consider a scattering problem defined by Eq.(\ref{SE}) on
a semi-infinite ($n\geq 0$) chain with no disorder
($\varepsilon_{n} = 0$) at sites
$n \geq N + 1$ and the boundary condition $\psi_{n=0}=0$. At $n \geq N $,
the wave function is taken in the form of an incident and reflected waves
\begin{equation}
\label{scattering}
\psi_{n} = e^{-ik(n-N)} + r_{N}e^{ik(n-N)} \;\;\;\; ; \;\;\; r_{N}=\mathrm{e}^{i\theta_{N}}
\, ,
\end{equation}
the last equation is due to the unitarity. The choice of the ``hard wall"
boundary condition $\psi_{n=0}=0$ is not significant in the limit $N \rightarrow\infty$.
The ratios $\psi_{n+1}/\psi_{n} \equiv Y_{n}$ with $0 < n \leq N$
obey the recursive equation
\begin{equation}
\label{Riccati}
Y_{n} = E-\varepsilon_{n} - \frac{1}{Y_{n-1}}
\,
\end{equation}
with the boundary condition $1/Y_{n=0} \equiv 0$. It is crucial that $Y_{n}$
knows only about $Y_{m}$
at $m\leq n$ and is insensitive to the distribution of
$\varepsilon_{m}$ at $m>n$. That is why the quantities
$Y_n$ at $0 < n \leq N$ coincide with the corresponding quantities
$\psi^{<}_{n+1}/\psi^{<}_{n}$ for semi-infinite chain in the Lyapunov problem
Eqs.(\ref{gele}) and (\ref{phi-def}):
\begin{equation}\label{Riccati-new}
Y_{n} = \frac{\psi^{<}_{n+1}}{\psi^{<}_{n}}=\frac{\cos(\phi_{n}+k)}{\cos(\phi_{n})}
\,  \;\;\;\; (0 < n \leq N) .
\end{equation}
As a consequence, the distribution of random $Y_{N}$ will be the
same as in the bulk  of a semi-infinite chain and thus governed by
the stationary distribution function ${\cal P}(\phi)$ of the
Lyapunov phase. On the other hand, it follows from
Eq.(\ref{scattering}) that
\begin{equation}\label{sc-anz}
Y_{N}=\frac{\psi_{N+1}}{\psi_N} = \frac{e^{-ik}+r_{N}\,e^{ik}}{1+r_{N}}=\frac{\cos(\frac{\theta_{N}}{2}+k)}{\cos(\frac{\theta_{N}}{2})}.
\end{equation}
Comparing Eqs.(\ref{Riccati-new}) and (\ref{sc-anz}) one concludes that
the distributions of the random quantities $\phi_{N}$ and $\theta_{N}$
are connected. At $N \rightarrow \infty$ the both distributions approach
their stationary limits, ${\cal P}(\phi)$ and $P_{{\rm ref}}(\theta)$,
respectively, with the relation:
\begin{equation}\label{theta-phi}
P_{{\rm ref}}(\theta)= \frac{1}{2}\,{\cal P}(\phi)|_{\phi = \theta/2}.
\end{equation}
Thus the distribution of the Lyapunov phase defined locally at each
link of the disordered chain  and the distribution of the global
reflection phase in a semi-infinite disordered chain  are related in
the simplest possible way.

\subsection{Exact relation between the generating function and the phase distribution function}
At vanishing disorder when the on-site energy distribution function
${\cal F}(\varepsilon)=\delta(\varepsilon)$, Eq.(\ref{Phi-equation})
equation reduces to
\begin{eqnarray}\label{Phi-ordered}
\Phi_{j+1}(z,\phi) = \mathrm{e}^{-z\cos^2{\phi}} \Phi_{j}(z, \phi -
k) \, .
\end{eqnarray}
In particular, at $z=0$ using the boundary conditions
Eq.(\ref{BC-z-phi}) we obtain:
\begin{eqnarray}\label{Phi-ordered-z-0}
\Phi_{j}(0,\phi) = \delta\left(\phi - j k-\frac{\pi}{2}\right) \, .
\end{eqnarray}
As the phase $\phi_{j}$ defined in Eq.(\ref{phi-def}) at vanishing
disorder is varying like $k j$ with the site number $j$,
Eq.(\ref{Phi-ordered-z-0}) suggests that $\Phi_{j}(z=0,\phi)$ is the
phase distribution function. Now we prove that this statement is
true at an arbitrary disorder.

Indeed, from definitions Eqs.(\ref{R-definition}) and
(\ref{W-definition}) we obtain
\begin{eqnarray}\label{W(0)}
\hspace{-2.5cm} W_j(s=0,v) = \left\langle
\exp{\left[-i\frac{v}{2}\left(\frac{1}{q^{+}_j} + \frac{1}{q^{-}_j}
\right)  \right]} \right\rangle  \Rightarrow  \tilde{W}_j(0,q)
=2\pi\left\langle \delta \left[ q -
\frac{\cos{(\phi_j)}}{\cos{(\phi_j +k)}} \right] \right\rangle ,
\end{eqnarray}
where it was taken into account the merging of $q^{+}_j$ and
$q^{-}_{j}$ in the limit $\eta \rightarrow 0$. Passing to the
variables $z$ and $\phi$, Eq.(\ref{z-q-variables}), we find for the
generating function $\Phi_{j}(z,\phi)$, Eq.(\ref{Phi-definition}),
at $z=0$:
\begin{eqnarray}\label{Phi(0)}
\Phi_j(z=0,\phi) = \left\langle \delta (\phi - \phi_j)\right\rangle
\equiv \mathcal{P}_{j}(\phi) \, .
\end{eqnarray}
Thus $\Phi_j(z=0,\phi)$ is equal to  the phase distribution function
$\mathcal{P}_{j}(\phi)$ corresponding to Eq.(\ref{phi-def}) for
solutions to Eq.(\ref{SE}) defined on a semi-infinite chain with a
generic boundary condition $\psi_{j=0} =0$, $\psi_{j=1} \sim 1$.
\begin{figure}[t]
\includegraphics[width=8cm, height=6
cm,angle=0]{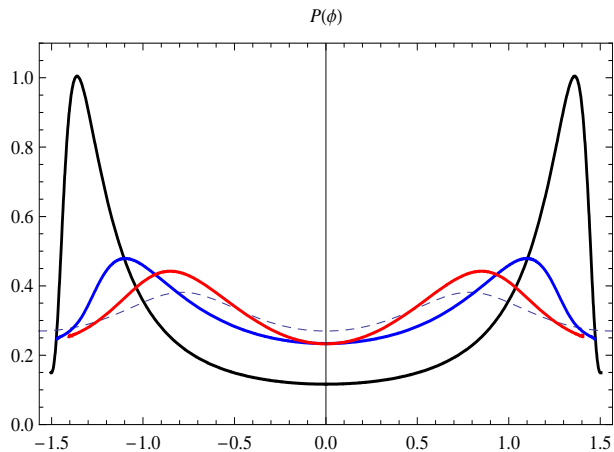} \caption{(color online) Stationary
distribution of the Lyapunov phase at
$E=0$ obtained from numerical solution of Eq.(\ref{P-equation}) for
Gaussian disorder with the dispersion $\sigma^{2}=0.1$ (red), 1.0
(blue), and 10 (black). The dotted line is the analytical solution
Eq.(\ref{P0-solution}) for $\sigma^{2}\rightarrow 0$ which is
$\pi/2$-periodic. With increasing disorder the maxima of the
distribution move towards $\phi=\pm \pi/2$ thus breaking the $\pi/2$
periodicity.}
\end{figure}

\begin{figure}[t]
\includegraphics[width=8cm, height=6
cm,angle=0]{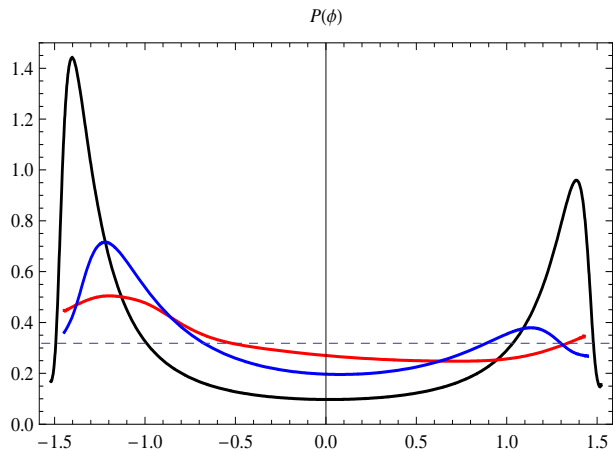} \caption{(color online) Stationary distribution of
the Lyapunov phase at
$k=1$ ($E=2 \cos(1)$) obtained from numerical solution of
Eq.(\ref{P-equation}) for Gaussian disorder with the dispersion
$\sigma^{2}=0.1$ (red), 1.0 (blue), and 10 (black). The dotted line
is the flat distribution ${\cal P}(\phi)=\frac{1}{\pi}$
corresponding to $\sigma^{2}\rightarrow 0$. Even at $\sigma^{2}=0.1$
(localization length $\ell_{0}\approx 80$) the distribution has a
pronounced structure and is far from being a constant ${\cal
P}(\phi)=\frac{1}{\pi}$.}
\end{figure}
The function $\mathcal{P}_{j}(\phi)$ is of interest in its own right
(see e.g. \cite{Wegner,Derrida}) as it determines the Lyapunov
exponent. This function obeys an exact recursive equation
\begin{eqnarray}\label{P-equation}
\mathcal{P}_{j+1}(\phi) =
\frac{\sin{k}}{ \cos^2{\phi}}
\int^{\pi}_{0} d\phi' {\cal F}\left((\sin k\,(\tan{\phi'} -
\tan{\phi})\right)\,
\mathcal{P}_{j}(\phi'-k ) \, ;\\
\mathcal{P}_{j=0}(\phi) = \delta(\phi -\pi/2) \, ,
\end{eqnarray}
which follows immediately from Eq.(\ref{Phi-equation}) for the
generating function $\Phi_j(z,\phi)$ at $z=0$. Eq.(\ref{P-equation})
is equivalent to the recursive Eq.(22) in Ref.\cite{Wegner}. For
weak disorder the stationary (site-independent) solution
$\mathcal{P}_{j}(\phi) \rightarrow \mathcal{P}(\phi)$ can be found
analytically ( see Ref.\cite{Derrida} and a brief discussion in
Sec.5). Examples of the distribution functions for different
strength of disorder and different energies obtained by
straightforward numerical solution of Eq.(\ref{P-equation}) are
presented in Fig.1 (for $E=0$ which corresponds to the  rational
$f=\frac{1}{2}$ with the smallest denominator) and Fig.2 (for the
irrational $f=\frac{1}{\pi}$).

The important relation Eq.(\ref{Phi(0)}) establishes an exact
correspondence between the generating function $\Phi_j(z=0,\phi)$
and the probability distribution function $\mathcal{P}_{j}(\phi)$
for the phase  $\phi_j$. This identity will be used later for the
proper normalization of the constructed ``stationary"
(site-independent) solution $\Phi(z,\phi)$. It shows also that the
generating function $\Phi_j(z,\phi)$ contains much more information
about the system than $\mathcal{P}_{j}(\phi)$. This is the reason
why the problem of our interest -- calculation of the statistical
moments of normalized eigenfunctions, Eqs.(\ref{IPR-m-definition}),
(\ref{IPR-m-integrals-z-phi-exact}), determined by the whole
$\Phi_j(z,\phi)$-- is much more difficult than calculation of the
Lyapunov exponent and similar quantities determined merely by
$\mathcal{P}_{j}(\phi)$. To emphasize the difference between the two
problems we note that the generalized IPR,
Eq.(\ref{IPR-m-integrals-z-phi-exact}), is not linear but
\emph{bi-linear} in $\Phi(z,\phi)$. Hence the generating function
$\Phi(z,\phi)$ itself \emph{cannot} be considered as a joint
probability distribution function of $z$ and $\phi$. The problem of
finding the joint probability distribution is not simple (see
section \ref{joint distribution}), and involves, as the first step,
the calculation of the moments $I_{m}$ for integer $m>0$. This will
be our main goal in this paper.

\subsection{Averaged density of states}\label{DOS-section}
For completeness, we apply the developed formalism to derive an expression for
the averaged local density of states (LDOS), determined by
\begin{eqnarray}\label{LDOS}
\nu(E,r) = -\frac{1}{\pi} \Im \, \langle G_{r,r}(E_{+})\rangle \, .
\end{eqnarray}
Using the representation Eq.(\ref{G-q}) for the retarded Green's function, definitions Eqs.(\ref{R-tilde-definition}), (\ref{R-tilde-definition}), and relation Eq.(\ref{R-tilda-R-identity}), we obtain
\begin{eqnarray}\label{LDOS-lambda}
\hspace{-2.cm}
\nu(E,r) = \frac{1}{\pi} \Im \left[i
\int^{\infty}_{0} d\lambda \, \mathrm{e}^{i\lambda E_{+}}
\left\langle\mathrm{e}^{-i\lambda \varepsilon_{r}}\right\rangle
\mathcal{R}_{r-1}(\lambda, 0)\mathcal{R}_{N-r}(\lambda, 0) \right]
\, .
\end{eqnarray}
Similar to the derivation of Eq.(\ref{IPR-m-integrals-z-phi-exact}), we use the definition Eq.(\ref{char}), relation Eq.(\ref{W-definition}): $\mathcal{R}_{j}(\lambda, 0) = W_{j}(\lambda\eta/2, \lambda) = \int \mathrm{e}^{-i\lambda q}\, \tilde{W}_{j}(\lambda\eta/2, q)\, dq/(2\pi)$, integrate over $\lambda$, and
take the limit $\eta \rightarrow 0$, to represent Eq.(\ref{LDOS-lambda}) in the form
\begin{eqnarray}\label{LDOS-lambda-q}
\hspace{-2.cm}
\nu(E,r) =
\int^{\infty}_{-\infty} \frac{d q \,d q'}{(2\pi)^2}\mathcal{F}(E-q-q')\tilde{W}_{r}(0, q) \tilde{W}_{N-r}(0, q')
\, .
\end{eqnarray}
In $(z, \,\phi)$ variables, Eq.(\ref{z-q-variables})-(\ref{Phi-definition}), the above expression
reads
\begin{eqnarray}\label{LDOS-z-phi}
&&\hspace{-2.cm}
\nu(E,r) =
\int^{\pi}_{0} d \phi \,d \phi' \, \mathcal{F}\left(E-\frac{\cos{\phi}}{\cos{(\phi + k)}} - \frac{\cos{\phi'}}{\cos{(\phi' + k)}}\right)\Phi_{r}(0, \phi) \Phi_{N-r}(0, \phi') \nonumber \\
&=& \int^{\pi}_{0} d \phi \,d \phi' \, \mathcal{F}\left(E-\frac{\cos{\phi}}{\cos{(\phi + k)}} - \frac{\cos{\phi'}}{\cos{(\phi' + k)}}\right)\mathcal{P}_{r}(\phi) \mathcal{P}_{N-r}(\phi')
\, .
\end{eqnarray}

\section{Weak disorder. Differential equation for the generating function}\label{PDE for GF}
An exact {\it integral} equation Eq.(\ref{Phi-equation}) contains
all the information about local statistics of eigenfunctions at any
uncorrelated on-site disorder. However, only at weak disorder the
statistical anomalies we are focusing at in this paper are sharp.
The point is that the region in the energy $E$ (or in the parameter
$k$) where statistics is anomalous is proportional to $\Delta E\sim
\sigma^{2}$ \cite{Derrida}, and for strong disorder $\sigma^{2}\sim
1$ the anomalies are rounded off. That is why in what follows we
consider only the case of weak disorder $\sigma^{2}\ll 1$.

For the case of a weak disorder, $\sigma^{2} \ll 1$, when the
``bare" (i.e.the one for the continuous model)) localization length
$\ell_{0} \gg 1$, the ``typical" squared amplitude of localized
eigenfunctions $z_{typ} \sim 1/\ell_{0} << 1$ and the exponential
factor in front of the integral in Eq.(\ref{Phi-equation}) can be
expanded in powers of $z$. From now we will introduce a re-scaled
variable
\begin{equation}\label{u-definition}
u \equiv \frac{1}{4}\,\ell_{0}\, z =
\frac{2\sin^{2}{(k)}}{\sigma^{2}}z \, ,
\end{equation}
keeping the notation $\Phi(u,\phi)=\Phi(z,\phi)|_{z=4u/\ell_{0}}$ for
the function of this variable.

\subsection{Expression for moments and DOS in case of weak disorder}
\label{weak disorder-moments-and-DOS} At weak disorder one can
replace ${\cal F}(\varepsilon)$ in
Eqs.(\ref{IPR-m-integrals-z-phi-exact}) and (\ref{LDOS-z-phi}) by a
$\delta$-function and perform integration over $\phi'$.  Using
Eq.(\ref{E-k}) one observes that vanishing of the argument in the
$\delta$-function results in
\begin{eqnarray}\label{transformation-phi}
\frac{\cos{\phi'}}{\cos{(\phi' + k)}}
= 2\cos{k} - \frac{\cos{\phi}}{\cos{(\phi + k)}} \Rightarrow -\tan{(\phi' +k)} = \tan{(\phi + k)} \, ,
\end{eqnarray}
from where it follows that
\begin{eqnarray}\label{phi}
 \phi' = -\phi - 2k \, , \,\,  \mathrm{mod}(\pi) \, .
\end{eqnarray}
As a consequence, $\cos^2{(\phi + k)}=\cos^2{(\phi' + k)}$, and we arrive at the following
expression for moments
\begin{eqnarray}\label{IPR-m-integrals-z-phi-weak}
I_{m}(r,E) &=&
\frac{4^{m-1}\,\nu_{0}(E)}{(m-2)!\,\nu(E)\,\ell^{m-1}_{0}}
\int^{\infty}_{0} d u \, \int^{\pi}_{0}
d\phi \,u^{m-2} \, \cos^{2m}({\phi}) \nonumber \\
&&\Phi_{r-1}(u,\phi-k)\, \Phi_{N-r}(u, -\phi -k) \, .
\end{eqnarray}
Here we have neglected  a small (at $\sigma^{2} \ll 1 \Rightarrow
\ell_{0} \gg 1$ ) exponent
$\mathrm{e}^{-4u\cos^{2}{(\phi+k)}/\ell_{0}}$ and used the freedom of
shifting the integration interval (as long as it equals to the
period of the integrand). In a similar way we find from
Eq.(\ref{LDOS-z-phi}) the averaged DOS in the weak disorder limit:
\begin{eqnarray}\label{LDOS-z-phi-weak}
&&\hspace{-2.cm}
\frac{\nu(E, r)}{\nu_{0}(E)} =
2\pi\int^{\pi}_{0} d \phi \, \cos^2{\phi} \, \mathcal{P}_{r}(\phi-k) \, \mathcal{P}_{N-r}(-\phi-k)
\, .
\end{eqnarray}
Up to notations, this relation between DOS and the phase probability distribution $\mathcal{P}_{r}(\phi)$
coincides with that derived in a different way in \cite{Wegner}.
For a long chain and far from its ends, the function $\mathcal{P}_{r}(\phi)$ becomes
site-independent and the averaged local DOS $\nu(E,r)$ coincides with $\nu(E)$ Eq.(\ref{DOS-definition}).

\subsection{Joint probability distribution function $P(u,\phi)$}\label{joint distribution}
By definition, the moments $I_{m}(r,E) \sim
N\langle|\psi|^{2m}\rangle$  are expressed via the probability
distribution function $P_{r}(|\psi|^2)$ as follows:
\begin{eqnarray}\label{IPR-P-psi}
I_{m}(r) = N\int^{\infty}_{0} |\psi|^{2m}\, P_{r}(|\psi|^2)\, d|\psi|^2
\, .
\end{eqnarray}
where the function $P_{r}(|\psi|^2)$ is yet to be found.
Starting from Eq.(\ref{IPR-m-integrals-z-phi-weak}), we first represent it in the form
\begin{eqnarray}\label{IPR-P-u-phi}
I_{m}(r) = N\int^{\infty}_{0} d u \, \int^{\pi}_{0}
d\phi \,[u \, \cos^{2}({\phi})]^{m} P_{r}(u, \phi) \, ,
\end{eqnarray}
which determines a ``joint probability distribution function" $P_{r}(u,\phi)$".
This function allows to find an average of an arbitrary function $f_{r}(u,\phi)$
\begin{eqnarray}\label{f-u-phi}
\langle f_{r}(u, \phi)\rangle = \int^{\infty}_{0} d u \, \int^{\pi}_{0}
d\phi \, f_{r}(u, \, \phi) P_{r}(u, \phi) \, ,
\end{eqnarray}
thus providing a more detailed information as compared to $P_{r}(|\psi|^2)$.
The connection between the two distribution functions is given by:
\begin{eqnarray}\label{P-psi-P-u-phi}
\hspace{-2.cm}
P_{r}(|\psi|^2) = \int^{\infty}_{0} d u \, \int^{\pi}_{0}
d\phi \, \delta(|\psi|^2 - u\cos^2{\phi}) P_{r}(u, \phi) = \int^{\pi}_{0}
\frac{d\phi}{\cos^2{\phi}}P_{r}\left(\frac{|\psi|^2}{\cos^2{\phi}}\, , \phi\right)\, .
\end{eqnarray}
Now, we will show that the joint probability distribution function $P_{r}(u,\phi)$
can be expressed via the basic object of our study, the generating function $\Phi_{r}(u,\phi)$.
Using in Eq.(\ref{IPR-m-integrals-z-phi-weak}) the representation
\begin{equation}\label{z-(m-2)-representation}
\frac{u^{m-2}}{(m-2)!} = \frac{d}{du}u^{m-1}\int_{-i\infty+0}^{+i\infty+0} \frac{d t}{2\pi i} \,\,\frac{e^{t}}{t^m} \, ,
\end{equation}
integrating in $u$ by parts, and changing $u \rightarrow ut$, we
represent Eq.(\ref{IPR-m-integrals-z-phi-weak}) in the form
Eq.(\ref{IPR-P-u-phi}), where $P_{r}(u,\phi)$ is given by
\begin{eqnarray}\label{P-u-phi-answer}
\hspace{-2.cm} P_{r}(u,\phi)= -\frac{\nu_{0}(E)}{N\nu(E)
u}\,\partial_{u}\,\int_{-i\infty+0}^{+i\infty+0} \frac{dt}{2\pi
i}\,\frac{e^{4t/\ell_{0}}}{t}\, \Phi_{r-1}\left(u t,\phi -
k\right)\, \Phi_{N-r}\left(u t, -\phi - k\right) \, .
\end{eqnarray}

Eqs.(\ref{IPR-P-u-phi}) and (\ref{P-psi-P-u-phi}) suggest an
interpretation of the relation $|\psi|^2 = u \cos^2{\phi}$ as a
decomposition of a fast varying (from site to site) eigenfunction
$\psi_{\nu}(j) \sim \sqrt{u_j}\,\cos{(\phi_j + kj)}$ in terms of two
auxiliary slow variables $u_j$ and $\phi_j$. The joint distribution
function $P_{r}(u,\phi)$, Eq.(\ref{P-u-phi-answer}), allows one also to
study separate distribution functions of these variables,
\begin{eqnarray}\label{P-u-and-P-phi-definition}
P^{(u)}_{r}(u) = \int^{\pi}_{0}
d\phi \, P_{r}(u, \phi) \,\,\,\,\,\,\, \mathrm{and} \,\,\, \,\,\,
P^{(\phi)}_{r}(\phi) = \int^{\infty}_{0} d u \, P_{r}(u, \phi)  \, .
\end{eqnarray}
There is an interesting open question about a relation between the
just defined distribution function $P^{(\phi)}_{r}(\phi)$ of the {\it
eigenfunction phase} for a finite chain with the boundary condition $\psi =0$
at its both ends, and the distribution function of the Lyapunov
phase $\mathcal{P}_{r}(\phi)= \Phi_{r}(u=0, \phi)$ for a semi-infinite
chain. These questions go beyond the scope of the current paper.

What we want to emphasize again at the end of this section is that the complete
information about the system is conveniently encoded in the generating function
$\Phi_j(u,\phi)$. Our nearest task is to obtain and solve a
differential equation (in the weak disorder limit) for this
function.

\subsection{Recursive differential operator for the generating function}
For weak disorder ($\sigma^{2} \ll 1$), the function ${\cal
F}(\varepsilon)$ in Eq.(\ref{Phi-equation}) is strongly peaked at
$\varepsilon=0$ and the integration over $\phi'$ is effectively
restricted to a narrow vicinity of $\phi$. Expanding the remaining
part of the integrand in powers of $(\tan{\phi'}-\tan{\phi})$ and
keeping only the first order terms in $\sigma^{2}$, we represent
Eq.(\ref{Phi-equation}) in the form:
\begin{equation}\label{diff-recursive-equation}
\Phi_{j+1}(u,\phi)=\left[1+\frac{4}{\ell_{0}}\,
\left[{\cal L}(u,\phi)-c_{1}(\phi)\,u\right]\right]\,\Phi_{j}(u,\phi-k) \, ,
\end{equation}
where ${\cal L}(u,\phi)$ is the second order
differential operator
\begin{equation}\label{diff-operator}
\hspace{-1.cm}
{\cal L}(u,\phi)=c_{2}(\phi)\,u^{2}\partial^{2}_{u}+c_{3}(\phi)\,(u\partial_{u}-1)+c_{4}(\phi)\,
u\partial_{u}\partial_{\phi}+c_{5}(\phi)\,\partial_{\phi}+c_{6}(\phi)\,\partial^{2}_{\phi}\, ,
\end{equation}
The coefficients $c_{i}(\phi)$ in Eqs.(\ref{diff-recursive-equation}) and (\ref{diff-operator}) are all
combinations of $\cos(2\phi)$ and $\sin(2\phi)$ and at first glance do not show any nice structure:
\begin{eqnarray}\label{coefficients}
&& \hspace{-2.5cm} c_{1}(\phi)=\frac{1}{2}[1+\cos(2\phi)], \;\;\;\;
c_{2}(\phi)=1-\cos^{2}(2\phi)\nonumber \\
&& \hspace{-2.5cm}c_{3}(\phi)=-[1-\cos(2\phi)-2\cos^{2}(2\phi)]
,\;\;\;\;c_{6}(\phi)=\frac{[1+\cos(2\phi)]^{2}}{4}\nonumber \\ &&
\hspace{-2.5cm}\, \,c_{4}(\phi)=\sin(2\phi)[1+\cos(2\phi)],\;\;\;\;
c_{5}(\phi)=-\frac{3}{2}\sin(2\phi)[1+\cos(2\phi)].
\end{eqnarray}
  Note that in the leading order in the disorder strength
Eq.(\ref{diff-recursive-equation}) depends only on the {\it
variance} of the on-site disorder distribution
${\cal F}(\varepsilon)$ through the bare localization length $\ell_{0}$
given by Eq.(\ref{u-definition}).

 From Eq.(\ref{diff-recursive-equation}) and the
established relation Eq.(\ref{Phi(0)}), we can immediately write a
recursive equation for the phase distribution function
$\mathcal{P}(\phi)$:
\begin{equation}\label{diff-recursive-equation-P}
\mathcal{P}_{j+1}(\phi)=\left[1+\frac{4}{\ell_{0}}\,
{\cal L}(\phi)\right]\, \mathcal{P}_{j}(\phi-k) \, ,
\end{equation}
where
\begin{equation}\label{diff-operator-P}
{\cal L}(\phi)=-\frac{\partial}{\partial \phi}
\left[  \frac{\sin{(2\phi)}[1+\cos{(2\phi)}]}{2}  -
\frac{[1+\cos{(2\phi)}]^2}{4}\frac{\partial}{\partial \phi} \right]
\, .
\end{equation}
Equations (\ref{diff-recursive-equation}) and
(\ref{diff-recursive-equation-P}) are the functional rather than the
differential equations because of the different phase arguments in
the left- and right-hand sides.

\subsection{Differential equation at rational $k/\pi = m/n \neq 1/2$}
In Eqs.(\ref{diff-recursive-equation}) and
(\ref{diff-recursive-equation-P}) the phase argument experiences a
finite jump $-k$ at the transition from the site $j$ to $j+1$. When
$k = \pi m/n$ with a natural $m$ and $n$, then after $n$ transitions
the shift of the phase argument becomes multiple of $\pi$ and the
functional equations can be safely reduced (at weak disorder
$\ell_{0}\gg 1$) to the differential ones due to the periodicity of
$\Phi(u,\phi)$, Eq.(\ref{Phi-periodicity}). Thus, iterating
Eq.(\ref{diff-recursive-equation}) $n$ times we get a closed
equation for $\Phi_j(u,\phi) = \Phi_{j}(u,\phi-nk))$:
\begin{eqnarray}\label{n-iterations}
\hspace{-2.5cm} \Phi_{j+n}(u,\phi) =
&&\left[1+\frac{4}{\ell_{0}}\,\left[{\cal L}(u,\phi)
-c_{1}(\phi)\,u\right]\right]... \nonumber \\
&&\left[1+\frac{4}{\ell_{0}}\,\left[{\cal L
}(u,\phi-(n-1)k)-c_{1}(\phi-(n-1)k)\,u\right]\right]\,\Phi_{j}(u,\phi-nk)
\, .
\end{eqnarray}
Keeping  only first order terms in the disorder strength
$\sigma^{2}\sim \ell_{0}^{-1}$, we obtain:
\begin{eqnarray}\label{lin}
\hspace{-2.5cm} \Phi_{j+n}(u,\phi)-\Phi_{j}(u,\phi) =
\frac{4}{\ell_{0}} \left[\sum_{r=0}^{n-1}{\cal L}(\phi-r\,\pi
m/n)-u\sum_{r=0}^{m-1}c_{1}(\phi-r\,\pi m/n) \right]\Phi_{j}(u,\phi)
.
\end{eqnarray}
Here the result of  summation is extremely sensitive to the
particular value of $k = \pi m/n$ and this is the formal reason of
an emerging anomaly. Indeed, the functions Eq.(\ref{coefficients})
contain only terms $\sim 1$, $\mathrm{e}^{\pm 2i\phi}$, and
$\mathrm{e}^{\pm 4i\phi}$, for which we have
\begin{equation}\label{id1}
\sum_{r=0}^{n-1}e^{2i(\phi - r\,\pi m/n)}=0, \,\,\,\,\,\,\,\,\,\,\,
\sum_{r=0}^{n-1}e^{4i(\phi - r\,\pi m/n) } =
\left\{  \begin{array}{cc}
    0 & n > 2 \\
    2\mathrm{e}^{4i\phi} & n=2 \,\, .\\
  \end{array}
\right.
\end{equation}
Thus, for $k \neq \pi/2 $ (i.e., $E(k) \neq 0$), only
$\phi$-independent parts of the coefficients Eq.(\ref{coefficients})
survive in Eq.(\ref{lin}). Assuming $n \ll \ell_{0}$, expanding the
L.H.S. of Eq.(\ref{lin}), and introducing the ``continuous"
dimensionless coordinate $x=2j/\ell_{0}$ along the chain, we obtain:
\begin{equation}
\label{ord}
\partial_{x}\Phi(u,\phi)=\left[u^{2}\partial^{2}_{u}-u+\frac{3}{4}\partial^{2}_{\phi} \right]
\,\Phi(u,\phi) \, .
\end{equation}
The variables $u$ and $\phi$ are separated and one can immediately find a ``stationary" (i.e. independent of $x$) solution
\begin{equation}\label{ord-solution}
 \Phi(u,\phi)=\Phi(u) = \frac{2}{\pi}\sqrt{u}\,K_{1}(2\sqrt{u}) \, .
\end{equation}
This {\it zero mode} solution describes the limit of a long chain
with the length $N\gg \ell_{0}$, it is the only one which survives
at distances $x \gg \ell_{0}$. This solution has been earlier
obtained \cite{Kolok} in the continuous limit ($n \gg 1$). It also
arises in the theory of a multi-channel disordered wire
\cite{Efet-book,Mirlin2000}. As follows from Eq.(\ref{ord}),
non-zero modes decay at distances $x =2j/\ell_{0}\sim 1$ providing
so called ``phase randomization": the zero-mode solution corresponds
to the absolutely isotropic distribution of the phase $\phi$.

The corresponding moments $I_{m}$ ($m=1,2,...$) are found from
Eq.(\ref{IPR-m-integrals-z-phi-weak}) and are equal to:
\begin{equation}
\label{mom-stand} I^{norm}_{m}=\frac{(m-1)!}{(\ell_{0})^{m-1}}
\end{equation}

The solution Eq.(\ref{ord-solution}) corresponds to the following probability distribution of squared wave
functions $|\psi|^{2}$ ($=z\cos^2{(\phi)}$) in a long {\it strictly} one-dimensional system (amazingly, this result was
not known before):
\begin{equation}\label{1d-dist}
\mathcal{P}(|\psi|^{2})\, d\,|\psi|^{2}=\frac{
\ell_{0}}{N}\,\frac{{\rm
exp}\left(-|\psi|^{2}\ell_{0}\right)}{|\psi|^{2}}\, d \,|\psi|^{2},
\,\,\,\,\,\,\,(|\psi|^{2}\ell_{0} \gg e^{-N/\ell_{0}}).
\end{equation}
where $N$   is the chain length. Note that the distribution
Eq.(\ref{1d-dist}) is not normalizable, as the normalization
integral is logarithmically divergent \cite{rem1}. This divergency is an
artefact of the zero-mode approximation and is typical to
exponentially localized wavefunctions.  The point is that the
eigenfunction statistics changes drastically for very small values
of the amplitude $|\psi|^{2}\ell_{0} \ll e^{-N/\ell_{0}}$, where the
zero-mode approximation no longer applies.
 This is related with the fact
that the {\it envelope} (i.e. $|\psi|^{2}$ averaged over
oscillations) of the typical localized wave function cannot be
significantly smaller than $|\psi|^{2}\sim e^{-N/\ell_{0}}$. The
smaller values of the amplitude $|\psi|^{2}$ are due to oscillations
and nodes of wave functions which probability is different from that
of the envelope. In the exact, $N$-dependent distribution function
the logarithmic divergency of the normalization integral is cut at
$|\psi|^{2}\ell_{0} < e^{-N/\ell_{0}}$.

\section{Center-of-band anomaly, $k=\pi/2$}\label{Center-of-band-anomaly}
As is seen from Eq.(\ref{id1}) for $k=\pi m/n = \pi/2$, terms $\sim
\mathrm{e}^{\pm 4i\phi}$ survive in Eq.(\ref{lin}). This leads to a
drastic modification of the phase-isotropic equation Eq.(\ref{ord}):
\begin{eqnarray}\label{anomal-equation}
\hspace{-1.5cm}
\partial_{x}\Phi \equiv \left[\hat{L} - u \right]\Phi = &&\left[[1 - \cos{(4\phi)}]\,u^{2}\partial^{2}_{u} +
\sin{(4\phi)}\, u\partial_{u}\partial_{\phi} \right. \\ \nonumber &&
\left. \hspace{-2.5cm}  + \frac{3 +
\cos{(4\phi)}}{4}\partial^{2}_{\phi} + 2\cos{(4\phi)}\,u\partial_{u}
- \frac{3}{2}\sin{(4\phi)}\partial_{\phi} - 2\cos{(4\phi)} -
u\right] \Phi
 \, ,
\end{eqnarray}
where the differential operator $\hat{L}$ depends explicitly on
$\phi$. The variables $u$ and $\phi$ are not separable anymore,
which results in an emergent center-of-band ($k=\pi/2 \Rightarrow
E=0$) \emph{anomaly}: the generating function and the phase
distribution function become non-isotropic in $\phi$. The variables
$u$ and $\phi$ cannot be separated even for the stationary variant
of Eq.(\ref{anomal-equation}) describing the zero mode:
\begin{eqnarray}\label{anomal-equation-zero-mode}
[\hat{L} - u]\Phi(u,\phi) = 0  \, ,
\end{eqnarray}
Yet, due to a hidden symmetry of Eq.(\ref{anomal-equation}), a proper choice of coordinates allows to
separate variables in the stationary (zero mode) equation Eq.(\ref{anomal-equation-zero-mode}).
This will be done in the next subsection.

For completeness, we conclude this subsection by derivation of an
exact expression for the stationary distribution function of phase
$\mathcal{P}(\phi)=\Phi(u=0,\phi)$ (see Eq.(\ref{Phi(0)})). Taking
the limit $u\rightarrow 0$ in the stationary variant
Eq.(\ref{anomal-equation-zero-mode}) of Eq.(\ref{anomal-equation}),
we obtain the ordinary differential equation:
\begin{eqnarray}\label{anomal-equation-P}
    0 = && \left[ \frac{3 + \cos{(4\phi)}}{4}\partial^{2}_{\phi} - \frac{3}{2}\sin{(4\phi)}\partial_{\phi}
    - 2\cos{(4\phi)}\right] \mathcal{P}(\phi) = \nonumber \\
     && \partial_{\phi}\left[ \frac{3 + \cos{(4\phi)}}{4}\partial_{\phi} - \frac{1}{2}\sin{(4\phi)}\right]
     \mathcal{P}(\phi) \, .
\end{eqnarray}
The only periodic solution to this equation has the form \cite{Derrida}:
 \begin{eqnarray}\label{P0-solution}
\mathcal{P}^{(an)}(\phi) = \frac{4\sqrt{\pi}}{\Gamma^2(\frac{1}{4})} \, \frac{1}{\sqrt{3 + \cos{(4\phi)}}}
     \, ,
\end{eqnarray}
where the normalization constant provides the equality
$\int^{\pi}_{0} d\phi \, \mathcal{P}(\phi)  = 1$. Thus, the
distribution of the Lyapunov phases in a long weakly disordered
chain at the center of band ($k=\pi/2$) is not isotropic but has
maxima at $\phi=\pm\frac{\pi}{4}$. According to the interpretation
Eq.(\ref{links}) of the amplitude-phase variables, this implies the
tendency towards smaller difference between $|\psi_{j}|^{2}$ and
$|\psi_{j+1}|^{2}$, i.e. larger localization length. This phenomenon
has been coined as \emph{``the center-of-band-anomaly"}.

Now we proceed with our much more difficult task: solving not an
ordinary but the partial differential equation
(\ref{anomal-equation}) for the generating function $\Phi(u,\phi)$
of two variables.

\subsection{Hidden symmetry and separation of variables}
The integrability of the stationary equation (\ref{anomal-equation-zero-mode}) is shown in three steps. The step one is to pass from $(u, \phi)$ to a new set of variables $(u,v)$ with $v=u\cos{(2\phi)}$, and to introduce a new function
\begin{equation}\label{Phi-tilde}
\tilde{\Phi}(u,v) = \frac{1}{u}\Phi(u,\phi)|_{\cos{(2\phi)} = v/u} \, .
\end{equation}
In these variables the zero-mode equation Eq.(\ref{anomal-equation}) takes a very symmetric form:
\begin{eqnarray}
\label{stat-xy}
\sqrt{u^{2}-v^{2}}\,\left\{\partial_{u}\,\,\sqrt{u^{2}-v^{2}}\,\,\partial_{u}
+\partial_{v}\,\,\sqrt{u^{2}-v^{2}}\,\,\partial_{v}\right\}\,\tilde{\Phi} = \frac{u}{2}\,\,\tilde{\Phi} \, .
\end{eqnarray}
It is remarkable that the L.H.S. of this equation can be represented as $[D_{1}^{2}+D_{3}^{2}]\,\tilde{\Phi}$
where the operators $D_{1}$ and $D_{3}$ belong to the family of
three operators from the representation of the $sl_{2}$ algebra:
\begin{equation}\label{A-xy}
D_{1}= \sqrt{u^{2}-v^{2}}\,\,\partial_{u} \,\,\, ;
\,\,\, D_{2}=u\,\partial_{v}+v\,\partial_{u}\,\,\, ; \,\,\, D_{3}=-\sqrt{u^{2}-v^{2}}\,\,\partial_{v}
\end{equation}
with the commutation relations:
\begin{eqnarray}\label{algebra}
[D_{1},D_{2}]=-D_{3},\;  [D_{3},D_{1}]=D_{2},\;  [D_{2},D_{3}]=D_{1}.
\end{eqnarray}
Now it is clear that there is a hidden order in a set of
coefficients in Eq.(\ref{anomal-equation}) resulting from the
$SL(2)$ symmetry. The latter frequently manifests itself in various
scattering problems. However, Eq.(\ref{anomal-equation}) is
connected with even higher symmetry. Introducing a set of three
additional (mutually commuting) operators
\begin{eqnarray}\label{B-operators}
B_{1}= v,\;  B_{2} = \sqrt{u^{2}-v^{2}} ,\;  B_{3}=u \, ,
\end{eqnarray}
we can represent Eq.(\ref{anomal-equation}) in the form $[D_{1}^{2}+D_{3}^{2} - B_{3}/2]\,\tilde{\Phi} =0$.
The operators $D_{i}$ and $B_{i}$ constitute an algebra $D \bigoplus B$ with the commutative subalgebra $B$ and
commutation relations: $[D,D]= D$ (see Eq.(\ref{algebra})), $[B,B]=0$, and $[D,B] =B$; in more detail, the latter relation looks like: $[D_{i},B_{i}] = 0$ and
\begin{eqnarray}\label{algebra-extended}
&&[D_{1},B_{2}]=B_{3}\,\,\, ; \,\,\,   [D_{2},B_{1}]=B_{3} \,\,\, ; \,\,\,  [D_{3},B_{1}]=-B_{2}\,\,\, ; \nonumber \\
&&[D_{1},B_{3}]=B_{2} \,\,\, ; \,\,\,  [D_{2},B_{3}]=B_{1} \,\,\, ; \,\,\,  [D_{3},B_{2}]=B_{1} \, .
\end{eqnarray}
It is tempting to interpret $D \bigoplus B$ as the algebra of generators of rotations ($D$) and translations ($B$) of the 3d pseudo-euclidian space $R^{1,2}$. However, the question of a constructive application of this symmetry to the considered problem remains open. Below we follow a more prosaic way.

The next step is to introduce a function
\begin{equation}\label{Psi-definition}
\Psi(u,v)=(u^{2}-v^{2})^{\frac{1}{4}}\,\tilde{\Phi}(u,v)
\end{equation}
to transform Eq.(\ref{stat-xy}) to the Schr\"{o}dinger-like equation for the function $\Psi(u,v)$:
\begin{eqnarray} \label{SE-Psi}
H\Psi\equiv
-(\partial_{u}^{2}+\partial_{v}^{2})\,\Psi+U(u,v)\,\Psi=0, \\
U(u,v)=-\frac{3}{4}\,\frac{u^{2}+v^{2}}{(u^{2}-v^{2})^{2}}+\frac{1}{2}\,
\frac{u}{u^{2}-v^{2}} \,.      \label{U}
\end{eqnarray}
Finally we introduce the variables
\begin{equation}\label{xi-eta}
\xi=\frac{u+v}{2}=u\,\cos^{2}{\phi} \, ,\;\;\;\; \eta=\frac{u-v}{2}=u\,\sin^{2}{\phi} \, .
\end{equation}
It is easy to see that in these  variables the operator in
Eq.(\ref{SE-Psi}) splits into two \emph{identical} {\it one-dimensional} Hamiltonians
\begin{eqnarray} \label{SE-xi-eta}
[\hat{H}_{\xi}+\hat{H}_{\eta}]\Psi(\xi, \eta) =0 \, ,
\end{eqnarray}
where $\hat{H}_{\xi}$ is given by:
\begin{equation}\label{H-1d}
\hat{H}_{\xi}=-\partial_{\xi}^{2}-\frac{3}{16}\,\frac{1}{\xi^{2}}+\frac{1}{4\xi}.
\end{equation}
Thus, in new variables Eq.(\ref{xi-eta}) the partial differential equation  (\ref{anomal-equation-zero-mode})
for the generating function at $k=\pi/2$ is separable and can be reduced to the two ordinary
differential equations of the Schr\"{o}dinger type
\begin{equation}\label{SE-two-ordinary}
\hat{H}_{\xi}\psi_{\Lambda}(\xi)=\Lambda\psi_{\Lambda}(\xi) \,\,\, ; \,\,\,
\hat{H}_{\eta}\psi_{-\Lambda}(\eta)=-\Lambda\psi_{-\Lambda}(\eta) \, ,
\end{equation}
defined on semi-axes $\xi \geq 0$ and $\eta \geq 0$, respectively.
The opposite sign of the two eigenvalues guarantees the zero-energy solution to Eq.(\ref{SE-xi-eta}).

\subsubsection{Distinctions from the usual quantum mechanics.}
Although Eqs.(\ref{SE-two-ordinary}) look like ordinary
one-dimensional Schr\"{o}dinger equations on a positive semi-axis,
the problem we are solving is very different from quantum mechanics.
A cornerstone of the latter is the Hermiticity of a Hamiltonian which insures
that corresponding eigenenergies are real and the time evolution of an
initial state is unitary. For singular Hamiltonians like the one in
Eq.(\ref{SE-two-ordinary}), this property is not given for granted.
It requires vanishing the boundary term
\begin{eqnarray}\label{herm}
&&\int^{\infty}_{0} d\xi\,\psi^{*}_{1}(\xi)\,\hat{H}\,\psi_{2}(\xi)
- \int^{\infty}_{0} d\xi\,\psi_{2}(\xi)\,\hat{H}\,\psi^{*}_{1}(\xi) \nonumber \\
&& = \left[\psi^{*}_{1}(\xi)\partial_{\xi}\psi_{2}(\xi)-
\psi_{2}(\xi)\partial_{\xi}\psi^{*}_{1}(\xi)\right]^{\xi=\infty}_{\xi=0}=
0 \, ,
\end{eqnarray}
which arises at integration by parts for any two quadratically
integrable functions $\psi_{1}(\xi)$ and $\psi_{2}(\xi)$ from the
Hilbert space. We will show that this condition cannot be fulfilled
for the operator Eq.(\ref{H-1d}) which eigenfunctions at $\Lambda
\neq 0$ are constructed as a superposition of two fundamental
solutions with different behaviors at $\xi \rightarrow 0^{+}$
\begin{equation}\label{fundamental-solutions superposition}
\psi_{-}(\xi) \sim \xi^{s_{-}}[1 + O(\xi)] \,\,\,\,\,\, ; \,\,\,\,\,\,
\psi_{+}(\xi) \sim \xi^{s_{+}}[1 + O(\xi)] \,.
\end{equation}
Here the exponents $s_{-}= 1/4$ and $s_{+}=3/4$ are the roots of the secular equation $s(s-1) + \frac{3}{16} = 0$.
For $\Lambda > 0$ the both solutions oscillate at $\xi \rightarrow \infty$
and are acceptable. For $\Lambda < 0$ the both solutions have exponentially decreasing and
increasing parts at $\xi \rightarrow \infty$, therefore only a properly constructed superposition
of the two solutions, with cancelation of the increasing part, is acceptable.
In the special case $\Lambda = 0$, the fundamental solution is
\begin{equation}\label{special-solution-1d}
\psi_{0}(\xi) \sim \xi^{1/4}\exp{\left(-\sqrt{\xi}\right)} \, ,
\end{equation}
while the second solution, $\sim \xi^{1/4}\exp{\left(-\sqrt{\xi}\right)}$, is unbounded and thus should be ignored.
Now, considering the equation Eq.(\ref{herm}) with the choice $\psi_{2}(\xi) \sim \psi_{+}(\xi)$ (for any $\Lambda > 0$)
and $\psi_{1}(\xi) \sim \psi_{0}(\xi)$, we find that the
boundary term vanishes at $\xi \rightarrow \infty$ while it is not zero at $\xi \rightarrow 0$:
\begin{equation}\label{boundary term}
\sim \left[\psi^{*}_{0}(\xi)\partial_{\xi}\psi_{+}(\xi)-
\psi_{+}(\xi)\partial_{\xi}\psi^{*}_{0}(\xi)\right]_{\xi=0} = \frac{1}{2} \neq 0 \, .
\end{equation}
This means the Hamiltonian Eq.(\ref{H-1d}) is \emph{non-Hermitian}
in the Hilbert space which includes all the (bounded) fundamental
solutions. The remedy of the usual quantum mechanics
\cite{Landau-QM} to preserve the Hermiticity of singular
Hamiltonians like Eq.(\ref{H-1d}) is to reduce the Hilbert space to
include only one of the two types (characterizing by the exponents
$s_{\pm}$) fundamental solutions. A direct consequence of such a
restriction is the absence of the bound states of negative energy
\cite{Landau-QM}. Indeed, as mentioned above, for $\Lambda <0$ only
a superposition of two fundamental solutions can provide a decrease
of the wave function at $\xi \rightarrow \infty$.

There is no such a restriction imposed on the Hilbert space in our problem
where the eigenvalue $\Lambda$ plays an auxiliary role (the zero mode of Eq.(\ref{SE-xi-eta})
arises as a result of cancelation $\Lambda+(-\Lambda)=0$) and does not have meaning
of an observable. That is why $\Lambda$ is allowed to be complex and there is no requirement
of Hermiticity of the operator Eq.(\ref{H-1d}). We will see that in contrast to quantum mechanics, bound states (decaying at $\xi \rightarrow \infty$) do exist in our problem.  Moreover, their spectrum is not discrete but fills a sector on the complex $\Lambda$ plane.

\subsubsection{Construction of a general solution, $\Lambda \neq 0$.}
Solutions $\psi_{\Lambda}(\xi)$ with $\Lambda \neq 0$ can be expressed via the Whittaker
function $W_{-\lambda, \, \mu}(\xi)$ which obeys the Weber's
differential equation (see, e.g. Ref. \cite{GR}, 9.220):
\begin{equation}\label{Weber-equation}
 \frac{d^2}{dx^2} \, W_{-\lambda,\,\mu}(x) + \left(-\frac{1}{4} - \frac{\lambda}{x} +
 \frac{\frac{1}{4} - \mu^2}{x^2} \right)W_{-\lambda,\,\mu}(x) = 0 \,
\end{equation}
and decays at $x \rightarrow \infty$:
\begin{eqnarray}\label{Whittaker}
&&W_{-\lambda, \, \mu}(x) = \frac{\Gamma(-2\mu)}{\Gamma(\frac{1}{2} - \mu + \lambda)}
M_{-\lambda, \, \mu}(x) +
\frac{\Gamma(2\mu)}{\Gamma(\frac{1}{2} + \mu + \lambda)}M_{-\lambda, \, -\mu}(x) \, ;\\
&&M_{-\lambda, \, \mu}(x) = x^{\frac{1}{2}+\mu}\mathrm{e}^{-x/2}
{}_{1}F_{1}(\frac{1}{2} + \mu + \lambda, 2\mu + 1; x) \, , \label{M-function-plus} \\
&&M_{-\lambda, \, -\mu}(x) = x^{\frac{1}{2}-\mu}\mathrm{e}^{-x/2}
{}_{1}F_{1}(\frac{1}{2} - \mu + \lambda, -2\mu + 1; x) \label{M-function-minus}
\, ,
\end{eqnarray}
where $_{1}F_{1}$ is the confluent hypergeometric function, and $ \Re \, \lambda \geq 0$.
For $\Lambda < 0$, the equation (\ref{H-1d}) is mapped on
Eq.(\ref{Weber-equation}) by the following identification:
\begin{equation}\label{identification}
\lambda = \frac{1}{8\sqrt{-\Lambda}}  \,\,\,\, ; \, \,\,\, x = 2\sqrt{-\Lambda}\,\xi = \frac{\xi}{4\lambda}\,\,\, \, ; \,\,\,\, \mu =\frac{1}{4}
 \, .
\end{equation}
Note that the solution $\psi_{\Lambda}(\xi) = W_{-\lambda, \,
\frac{1}{4}}\left(\frac{\xi}{4\lambda}\right) $ to
Eqs.(\ref{SE-two-ordinary}),(\ref{Weber-equation}) at $\Lambda<0$,
which decays at $\xi\rightarrow\infty$, contains both a part $\sim
\xi^{\frac{1}{4}}$ and a part $\sim \xi^{\frac{3}{4}}$ at
$\xi\rightarrow 0$.  This clearly violates the condition of
Hermiticity Eq.(\ref{herm}) and is the reason why the singular
Hamiltonian Eq.(\ref{H-1d}) does have bound states.
Moreover, the spectrum of the Hamiltonian is
complex, as Eq.(\ref{identification}) can be easily extended to complex
$\Lambda$ with the convention that $\sqrt{z}>0$ at $z>0$ and has a
cut along the semi-axis $z<0$. From the asymptotic of the Whittaker function
\begin{equation}\label{asymptotic}
W_{-\lambda, \, \frac{1}{4}}\left(\frac{\xi}{4\lambda}\right) \sim \left(\frac{\xi}{4\lambda}\right)^{-\lambda} \exp{\left(-\frac{\xi}{8\lambda}\right)} \, \,\, \, ; \,\,\,\, \xi \rightarrow \infty \, ,
\end{equation}
we find the domain on the complex plane of $\lambda = |\lambda|\mathrm{e}^{i\alpha}$, where the
solution Eq.(\ref{asymptotic}) is decaying: $\Re \lambda > 0$, i.e., $\alpha \in (-\pi/2, \pi/2)$.
A general solution to Eq.(\ref{SE-Psi}) can be built as a superposition of "elementary blocks"
\begin{equation}\label{product}
W_{-\lambda_1, \, \frac{1}{4}}\left(\frac{\xi}{4\lambda_1}\right) \,\, W_{-\lambda_2, \, \frac{1}{4}}\left(\frac{\eta}{4\lambda_2}\right) \, ,
\end{equation}
with the following restrictions for $\lambda_1$ and $\lambda_2$: (i) $\lambda^2_1 + \lambda^2_2 = 1/(-\Lambda) + 1/\Lambda = 0$ ; (ii) $\Re \lambda_1 \geq 0$, $\Re \lambda_2 \geq 0$. These inequalities are non-strict because in the case where one of the $\lambda$'s is purely imaginary the second one is real due to the condition (i) and this provides the required vanishing of the product Eq.(\ref{product}) at $u \rightarrow \infty$. The relation (i) leads to the representation $\lambda_1 = |\lambda|\mathrm{e}^{i\alpha}$, $\lambda_2 = |\lambda|\mathrm{e}^{i\beta}$ with $\alpha, \beta \in [-\pi/2, \pi/2]$ and $\beta = \alpha \pm \pi/2$. Choice of the upper (lower) sign means that $\lambda_1$ is restricted to the IV (I) quadrant of the complex plane with $\alpha \in [-\pi/2, 0]$ ($\alpha \in [0, \pi/2]$).

As a result, the general solution  to Eq.(\ref{SE-Psi}) can be
represented as a superposition
\begin{eqnarray}\label{superposition-lambda}
&&\Psi(\xi, \eta) = \int_{ \Re \, \lambda \geq 0, \Im \, \lambda \geq 0} d^2\lambda \,\,
c_{+}(\lambda,\bar{\lambda})\,\,W_{-\lambda, \,
\frac{1}{4}}\left(\frac{\xi}{4\lambda}\right) \,\, W_{i\lambda, \, \frac{1}{4}}\left(\frac{i\eta}{4\lambda}\right) \nonumber \\
&& \hspace{-2cm} + \int_{ \Re \, \lambda \geq 0, \Im \, \lambda \leq 0} d^2\lambda \,\,
c_{-}(\lambda,\bar{\lambda})\,\,W_{-\lambda, \,
\frac{1}{4}}\left(\frac{\xi}{4\lambda}\right) \,\, W_{-i\lambda, \, \frac{1}{4}}\left(\frac{-i\eta}{4\lambda}\right) \equiv \Psi_{+}(\xi, \eta) + \Psi_{-}(\xi, \eta) \, ,
\end{eqnarray}
where $c_{\pm}(\lambda,\bar{\lambda})$ are arbitrary functions (restricted
by the requirement of convergency of the corresponding integrals).

We mention for completeness, that the Whittaker functions with the
second index $\mu = 1/4$ which emerges in our problem, constitute a
special class. They can be expressed in terms of the \emph{parabolic
cylinder functions} $D_{\kappa}(x)$: $W_{\lambda, \, \frac{1}{4}}(x)
= 2^{-\lambda}(2x)^{1/4}D_{2\lambda - \frac{1}{2}}(\sqrt{2x})$,
which reflects the possibility of mapping the problem
Eq.(\ref{SE-two-ordinary}) on the problem of two ``harmonic
oscillators" but with an ``upside down'' potential for one of them. We
will not explore this correspondence in the present paper.

\subsubsection{Solution in the case $\Lambda = 0$.} For the eigenvalue $\Lambda = 0$,
the mapping Eq.(\ref{identification}) becomes singular. Two eigenfunctions of the operator Eq.(\ref{H-1d})
in this simple case are elementary functions, increasing and decreasing at $\xi \rightarrow \infty$, respectively.
The eigenfunction $\psi_0(\xi)$ decreasing at $\xi \rightarrow \infty$ is given by Eq.(\ref{special-solution-1d}).
However, as will be shown in the next section, the corresponding solution of Eq.(\ref{SE-two-ordinary})
\begin{equation}\label{special-solution-2d}
\Psi_{0}(\xi,\eta) = (\xi\eta)^{1/4} \exp{\left[-\left(\sqrt{\xi} + \sqrt{\eta}\right)\right]}  \,
\end{equation}
does not meet physical requirements of smoothness of $\Phi(u,\phi)$
as a function of $\phi$ and thus it must be ignored.

\subsubsection{Problem of degeneracy of the general solution.}
Note that Eq.(\ref{superposition-lambda}) possesses a huge
degeneracy, due to an arbitrary choice of the functions
$c_{\pm}(\lambda,\bar{\lambda})$. This is in contradiction with an
intuitive expectation that the statistics of wave functions in an
infinite disordered chain should be unique and independent of the
boundary conditions. Below we show that the natural physical
requirement of smoothness of $\Phi(u,\phi)$ as a function of $\phi$
helps to determine the solution for the generating function up to a
constant pre-factor which can be further fixed using the relation
Eq.(\ref{Phi(0)}) and the normalization condition
$\int^{\pi}_{0}\mathcal{P}^{(an)}(\phi) d\phi =1$ for the anomalous
phase distribution function $\mathcal{P}^{(an)}(\phi)$.

\section{Resolving the degeneracy problem and determining the solution for the generating function}\label{Solution for GF}
\subsection{Requirements for the generating function $\Phi(u,\phi)$ } \label{requirements}
The stationary generating function
\begin{eqnarray}\label{Phi-vs-Psi}
\hspace{-1.5cm}
\Phi(u,\phi) \equiv \left\{\Phi(\xi,\eta)\right\}_{\xi=u\cos^{2}{\phi} \, , \, \eta=u\sin^{2}{\phi}} =
\left\{\frac{\xi + \eta}{(\xi\eta)^{1/4}}\Psi(\xi,\eta)\right\}_{\xi=u\cos^{2}{\phi} \, \,\, , \,\,\, \eta=u\sin^{2}{\phi}}
 \, ,
\end{eqnarray}
which determines the moments Eq.(\ref{IPR-m-integrals-z-phi-weak}),
should obey the following requirements:

1. It should vanish at $u \rightarrow \infty$.

2. It should be periodic in $\phi$ (with the period $\pi$). Moreover, for the considered case ($k=\pi/2$) the
coefficients of Eq.(\ref{anomal-equation}) are periodic functions with the period $\pi/2$ and we impose the
requirement $\Phi(u, \phi + \pi/2) = \Phi(u, \phi)$ on the stationary solution, too.

3. It should obey the relation $\Phi(u = 0,\phi) = \mathcal{P}^{(an)}(\phi)$, which follows from
Eqs.(\ref{Phi(0)}) and (\ref{P0-solution}).

4. $\Phi(u,\phi)$ should be a smooth function of $\phi$ together with all the derivatives with respect to $\phi$.
It should have no jumps, cusps, etc.

The first requirement has been fulfilled due to the proper choice of the integration domains in
Eq.(\ref{superposition-lambda}) on the complex plane $\lambda$.
The second requirement is equivalent to $\Psi(\xi, \eta)
= \Psi(\eta, \xi)$ and can be achieved by the symmetrization of the
integrand in Eq.(\ref{superposition-lambda}) with respect to the
replacement $\xi \leftrightarrow \eta$. The third requirement will be fulfilled by the properly
chosen behavior of the functions $c_{\pm}(\lambda,\bar{\lambda})$ at $|\lambda| \rightarrow 0$
(see subsection \ref{Equation for C} and Appendix A).
Accounting for the fourth, extremely important requirement is not as simple. It will be
postponed till subsections \ref{Equation for C}, \ref{Solution for C}.

\subsection{From plane to contour integral}\label{subs-contour}
Similarly to the usual coherent states, the set of partial solutions
Eq.(\ref{product}) is overcomplete, and actually the 2d integration
domains in Eq.(\ref{superposition-lambda}) (i.e., the first and the fourth
quadrants of the complex plane $\lambda$) can be reduced without losses to a
1d integration contour. Consider, for instance, the first term in
Eq.(\ref{superposition-lambda}). Note that $F_{+}(\lambda;\xi,\eta)
\equiv W_{-\lambda, \, \frac{1}{4}}\left(\frac{\xi}{4\lambda}\right)
\,\, W_{i\lambda, \,
\frac{1}{4}}\left(\frac{i\eta}{4\lambda}\right)$ is a
\emph{holomorphic function} of $\lambda$ in the first quadrant,
i.e., $F_{+}(\lambda;\xi,\eta)$ depends only on $\lambda=\rho
e^{i\alpha}$ ($\rho \equiv |\lambda|$) but not on $\bar{\lambda}=\rho e^{-i\alpha}$. In polar
coordinates $(\rho, \alpha)$ we have:
\begin{eqnarray}\label{Psi-polar}
&&\Psi_{+}(\xi, \eta) = \int^{\pi/2}_{0} d\alpha
\int_{\Gamma_0} \rho \, d\rho \, c_{+}(\rho, \alpha) \, F_{+}(\rho
\mathrm{e}^{i\alpha}; \xi, \eta) = \nonumber \\
&&\int^{\pi/2}_{0} d\alpha
\int_{\Gamma_{\alpha}} \rho \, d\rho \, c_{+}(\rho,
\alpha)\,F_{+}(\rho\mathrm{e}^{i\alpha}; \xi, \eta) \,
\end{eqnarray}
where the integration contour $\Gamma_0$ coincides with the
semi-axis $\lambda \geq 0$, and the second equality is the
realization of the possibility to rotate the contour $\Gamma_0$,
provided that the first argument $\rho \mathrm{e}^{i\alpha}$ of the
function $F_{+}$ remains in the first quadrant. For instance, if we
choose the contour $\Gamma_{\alpha}$ as a ray which corresponds to
the rotation of $\Gamma_0$ by the angle $-\alpha$, the variable
$\rho$ on $\Gamma_{\alpha}$ is represented as
$|\rho|\mathrm{e}^{-i\alpha}$. Changing, at a fixed $\alpha$, the
variable $\rho$ in the internal integral: $\rho =
t\mathrm{e}^{-i\alpha}$, where real $t$ runs from $0$ to $+\infty$,
we arrive at:
\begin{eqnarray}\label{Phi-t}
\hspace{-2.cm}
\Psi_{+}(\xi, \eta) = \int^{\pi/2}_{0} d\alpha \int^{+\infty}_{0} dt \, \mathrm{e}^{-2i\alpha} \,t
\, c_{+}(t\mathrm{e}^{-i\alpha}, \alpha) \, F_{+}(t; \xi, \eta)
\, ,
\end{eqnarray}
where we have changed the order of integrations, introduced a new weight function
\begin{eqnarray}\label{C-new}
\mathcal{C}_{+}(t) \equiv t\int^{\pi/2}_{0} d \alpha \, \mathrm{e}^{-2i\alpha} \, c(t\mathrm{e}^{-i\alpha}, \alpha)
\, ,
\end{eqnarray}
and changed back the notation $t \rightarrow \lambda$. Thus without
loss of generality we have expressed the double integral over the first quadrant in terms of a
\emph{contour} integral
\begin{eqnarray}\label{I-polar}
\Psi_{+}(\xi, \eta) = \int_{\Gamma_0} d\lambda \, C_{+}(\lambda) \,
F(\lambda;\xi,\eta) \, .
\end{eqnarray}
Note again, that the key condition for this transformation is a
holomorphic dependence of $F_{+}(\lambda;\xi,\eta)$ on $\lambda$.

Another choice of the integration contour can make the
expression more symmetric. Namely, rotating the contour $\Gamma_0$ by the
angle $\pi/4$, so that $\lambda \rightarrow
|\lambda|\mathrm{e}^{i\pi/4}$, and introducing a new real variable
$\lambda'$ by $\lambda = \lambda'\mathrm{e}^{i\pi/4}$, we obtain
(omitting the prime and re-defining the arbitrary function
$C_{+}(\lambda)$)
\begin{eqnarray}\label{canon-plus}
\Psi_{+}(\xi,\eta)=\int_{0}^{\infty}d\lambda\,
C_{+}(\lambda)\, W_{-\lambda\epsilon,\frac{1}{4}}\,\left(
\frac{\bar{\epsilon}\xi}{4\lambda}\right)W_{-\lambda\bar{\epsilon},\frac{1}{4}}\,\left(
\frac{\epsilon\eta}{4\lambda}\right)\, .
\end{eqnarray}
Here $\epsilon=e^{i\pi/4}$, $\bar{\epsilon}=e^{-i\pi/4}$.

Following the same route we obtain a contour integral representation for the second term in
Eq.(\ref{superposition-lambda}):
\begin{eqnarray}\label{canon-minus}
\Psi_{-}(\xi,\eta)=\int_{0}^{\infty}d\lambda\,
C_{-}(\lambda)\, W_{-\lambda\bar{\epsilon},\frac{1}{4}}\,\left(
\frac{\epsilon\xi}{4\lambda}\right)W_{-\lambda\epsilon,\frac{1}{4}}\,\left(
\frac{\bar{\epsilon}\eta}{4\lambda}\right)\,
\end{eqnarray}
with $C_{-}(\lambda)$ related with $c_{-}(\lambda, \bar{\lambda})$.

Eqs.(\ref{canon-plus}) and (\ref{canon-minus}) determine the generating function $\Phi(u, \phi)$
Eq.(\ref{Phi-vs-Psi}):
\begin{eqnarray}\label{canon}
\Phi(\xi,\eta)=\frac{\xi+\eta}{(\xi\eta)^{1/4}}\int_{0}^{\infty}d\lambda\,
C(\lambda)\left[W_{-\lambda\epsilon,\frac{1}{4}}\,\left(
\frac{\bar{\epsilon}\xi}{4\lambda}\right)W_{-\lambda\bar{\epsilon},\frac{1}{4}}\,\left(
\frac{\epsilon\eta}{4\lambda}\right)+ c.c. \, \right] \, .
\end{eqnarray}
Here we took $C_{+}(\lambda)=C_{-}(\lambda) \equiv C(\lambda)$ to make the integrand
symmetric with respect to the permutation $\xi \leftrightarrow \eta$
in order to fulfill the formulated above requirement 2; the function $C(\lambda)$ is a real
(without loss of generality) function yet to be determined. Up to now we have used only the
following loose assumptions on its properties:

$1^o$. $C(\lambda)$ has no singularities in the first quadrant of the complex $\lambda$ plane;

$2^o$. at $|\lambda| \rightarrow \infty$, the integrand in Eq.(\ref{canon}) decays faster than $1/\lambda$;
this justifies rotations of the contour neglecting contributions of distant arcs.

\subsection{Equation for $C(\lambda)$}\label{Equation for C}
We begin by determining the behavior of the function $C(\lambda)$ at
$\lambda\rightarrow 0$. To this end we note that according to the
relations Eqs.(\ref{Phi(0)}) and (\ref{P0-solution}), the generating
function $\Phi(\xi,\eta)$ Eq.(\ref{canon}) must tend to a finite
limit as $\xi\rightarrow 0$ and $\eta\rightarrow 0$.

Re-scaling in Eq.(\ref{canon}) the integration variable $\lambda \rightarrow u\lambda$, we find at $u \rightarrow 0$
\begin{eqnarray}\label{canon-u-0}
\hspace{-2.5cm} \Phi(0, \phi) \sim
\frac{u^{3/2}}{|\cos{\phi}\sin{\phi}|^{1/2}}\int_{0}^{\infty}d\lambda\,
C(u\lambda)\left[W_{0,\frac{1}{4}}\left(
\frac{\bar{\epsilon}\cos^2{\phi}}{4\lambda}\right)W_{0,\frac{1}{4}}\left(
\frac{\epsilon\sin^2{\phi}}{4\lambda}\right)+ c.c.\right].
\end{eqnarray}
To provide a finite value of the expression (\ref{canon-u-0}) in the
limit of vanishing $u$, we should require that (see Appendix A):
\begin{equation}\label{tile-non}
C(\lambda)=\frac{\tilde{C}(\lambda)}{\lambda^{\frac{3}{2}}}\;,\;\;\;\tilde{C}(0)={\rm const} \, .
\end{equation}
A crucial role in further restricting the possible choice of the
function $\tilde{C}(\lambda)$ is played by the requirement of
smoothness of $\Phi(u,\phi)$ as a function of $\phi$ (requirement 3
of the previous subsection). The generating function
Eq.(\ref{canon}) is periodic in $\phi$ with the period
$\frac{\pi}{2}$ and it is continuous at the end points $\phi=0$
(i.e. $\eta =0$), and $\phi = \pi/2$ (i.e. $\xi = 0$) of the
interval of the periodicity $(0,\pi/2)$. This is guaranteed by the
{\it c.c.} term in Eq.(\ref{canon}), equivalent to the permutation
$\xi \leftrightarrow \eta$. What is not automatically guaranteed is
that $\Phi(\xi,\eta)$ is {\it smooth} as a function of $\phi$ at
$\phi=0$ and $\phi = \pi/2$; the smoothness implies the continuity
of all the derivatives.  Amazingly, the requirement of {\it
smoothness} is sufficient to determine the function
$\tilde{C}(\lambda)$ up to a constant pre-factor. As we will see
this happens because of the special property of the solution
Eq.(\ref{canon}) encoded in the certain identity for the confluent
hypergeometric functions in
Eqs.(\ref{Whittaker})-(\ref{M-function-minus}).

Consider, for instance, the behavior of $\Phi(u,\phi)$ at $\phi
\rightarrow 0$, (i.e. $\eta \rightarrow 0$, while $\xi \rightarrow
u$). A discontinuity of derivatives at $\phi=0$ may arise from the
branching of the expression in Eq.(\ref{canon}) at small $\eta$.
Indeed, according to the representation of the Whittaker function
Eq.(\ref{Whittaker}) in terms of $M$-functions
Eqs.(\ref{M-function-plus}) and (\ref{M-function-minus}), we see
that
\begin{eqnarray}\label{W-at-small-eta}
W_{-\lambda\bar{\epsilon},\frac{1}{4}}\,\left(\frac{\epsilon\eta}{4\lambda}\right) =
\left(\frac{\epsilon\eta}{4\lambda}\right)^{\frac{1}{4}}\left[f_1(\lambda, \eta) + \sqrt{\eta} \, f_2(\lambda, \eta)\right] \, ,
\end{eqnarray}
where $f_1$ and $f_2$ are analytic functions of $\eta$ in the
vicinity of $\eta = 0$. The common factor $\eta^{1/4}$ is canceled
by the pre-factor in front of the integral in Eq.(\ref{canon}). The
first term in the square brackets of Eq.(\ref{W-at-small-eta}) is
regular in the vicinity of $\eta=0$, while the second one $\sim
\sqrt{\eta} \sim |\phi|$, is not analytical at $\eta=0$. As such a
non-analytical behavior is in conflict with  the  requirement $3$ of
smoothness (section \ref{requirements}), the corresponding part of
the solution must identically
vanish. Extracting this singular ($\propto \sqrt{\eta}$ in the
domain $\eta < \xi$) part $\Phi_{sing}(\xi,\eta)$ of the general
solution Eq.(\ref{canon}), we obtain:
\begin{eqnarray}\label{C-equation}
\hspace{-2.cm} \Phi_{sing}(\xi,\eta) \sim \int^{+\infty}_{0}
d\lambda \, \frac{\tilde{C}(\lambda)}{\lambda^{3/2}} && \left[
W_{-\lambda\epsilon,\,\frac{1}{4}}\left(\frac{\bar{\epsilon}\xi}{4\lambda}\right)
M_{-\lambda\bar{\epsilon},\frac{1}{4}}\left(\frac{\epsilon\eta}{4\lambda}\right)\frac{1}{\Gamma(\frac{1}{4}
+ \bar{\epsilon}\lambda)}
\right. \nonumber \\
&& \left.
+ M_{-\lambda\epsilon,\,\frac{1}{4}}\left(\frac{\bar{\epsilon}\eta}{4\lambda}\right)
W_{-\lambda\bar{\epsilon},\,\frac{1}{4}}\left(\frac{\epsilon\xi}{4\lambda}\right)\frac{1}{\Gamma(\frac{1}{4}
+ \epsilon\lambda)}\right] = 0
 \, ,
\end{eqnarray}
which should be fulfilled for any $\eta < \xi$. Vanishing of
$\Phi_{sing}(\xi,\eta)$ is equivalent to the homogeneous integral
equation for the real weight function $\tilde{C}(\lambda)$ with the
boundary condition $\tilde{C}(\lambda \rightarrow 0) \rightarrow
\mathrm{const}$.

Similarly, the presence of non-analytical terms $\sqrt{\eta}$ and $\sqrt{\xi}$ in the special solution
Eq.(\ref{special-solution-2d}) is the reason why this solution should be ignored.

The integral equation Eq.(\ref{C-equation}) imposes severe
constraints on the function $\tilde{C}(\lambda)$, because
Eq.(\ref{C-equation}) must be satisfied \emph{for arbitrary} $\eta$
and $\xi$ (at $\eta < \xi$). That is why the requirement of
smoothness lifts a huge degeneracy and arbitrariness in the possible
choice of $\tilde{C}(\lambda)$. The existence of even a single
(non-zero) solution for $\tilde{C}(\lambda)$ is not evident. We will
show below that the solution to Eq.(\ref{C-equation}) does exist and
is unique up to the constant pre-factor.

\subsection{Solution for $\tilde{C}(\lambda)$}\label{Solution for C}
Rotating the integration contours independently for each of the two terms in the integrand of Eq.(\ref{C-equation}) and
changing $\lambda \rightarrow t\bar{\epsilon}$ and $\lambda \rightarrow t\epsilon$, respectively, one can make
the Whittaker function real and take it out of the square brackets. Thus, the integral equation Eq.(\ref{C-equation}) takes the form
\begin{eqnarray}\label{C-equation-factorized}
\int^{+\infty}_{0}\frac{dt}{t^{9/4}} \,W_{-\lambda,\frac{1}{4}}\left(\frac{\xi}{4t}\right)
&&\left[\frac{\tilde{C}(\bar{\epsilon}t)\exp{\left(-i\frac{\eta}{8t}\right)}}{\Gamma(\frac{1}{4} - it)}
{}_{1}F_{1}\left(\frac{3}{4}-it,\, \frac{3}{2}\,; \frac{i\eta}{4t}\right) \right. \nonumber \\
&& \left. - \frac{\tilde{C}(\epsilon t)\exp{\left(i\frac{\eta}{8t}\right)}}{\Gamma(\frac{1}{4} + it)}
{}_{1}F_{1}\left(\frac{3}{4}+it,\, \frac{3}{2}\,; \frac{-i\eta}{4t}\right) \right] = 0
 \, ,
\end{eqnarray}
where the dependence of the integrand on $\xi$ and $\eta$ is
factorized. The only possibility to satisfy this equation for
arbitrary $\xi$ and $\eta$ is to require the square bracket to
vanish identically.

The crucial observation for the possibility to fulfil this condition
is an identity for the confluent hypergeometric functions \cite{GR}:
\begin{equation}\label{gr-iden-part}
e^{-z/2}\,_{1}F_{1}\left(\frac{3}{4}-it,\frac{3}{2},z
\right)=e^{z/2}\,_{1}F_{1}\left(\frac{3}{4}+it,\frac{3}{2},-z
\right).
\end{equation}
With the help of Eq.(\ref{gr-iden-part}), we find that the square
bracket in Eq.(\ref{C-equation-factorized}) vanishes identically
\emph{for all} $\eta$ if and only if the function $\tilde{C}(t)$
obeys (for positive $t$) the condition
\begin{eqnarray}\label{C-final-equation}
\frac{\tilde{C}(\bar{\epsilon}t)}{\Gamma(\frac{1}{4} - it)} =
\frac{\tilde{C}(\epsilon t)}{\Gamma(\frac{1}{4} + it)}
 \, .
\end{eqnarray}
Now one can immediately guess \emph{a solution} for $\tilde{C}(\lambda)$:
\begin{equation}\label{C0-solution}
\tilde{C}_{0}(\lambda)=\Gamma\left(\frac{1}{4}+\epsilon\lambda
\right)\,\Gamma\left(\frac{1}{4}+\bar{\epsilon}\lambda \right) \, .
\end{equation}
It is easily seen that the solution Eq.(\ref{C0-solution}) obeys the both conditions formulated at
the end of the subsection \ref{subs-contour}. The function $C_{0}(\lambda) \equiv \tilde{C}_{0}(\lambda)/\lambda^{3/2}$
(see definition (\ref{tile-non})) is an analytical function in the domain of our interest ($\Re \,\lambda > 0$).
Though $C_{0}(\lambda)$ grows at $|\lambda| \rightarrow \infty$, one can check that the integrand in Eq.(\ref{canon})
decays as $1/|\lambda|^{3}$ for $|\lambda| \rightarrow \infty$. This provides the convergence of the integral
Eq.(\ref{canon}) at large $\lambda$ and justifies rotations of integration contours neglecting contributions
of infinitely remote arcs. Note also that $\tilde{C}_{0}(\lambda)$ is real at the semi-axis $\lambda > 0$.

Now, looking for \emph{a general} solution to Eq.(\ref{C-final-equation}) in the form
\begin{equation}
\label{gen-C}
\tilde{C}(\lambda)=C_{0}(\lambda)S(\lambda) \, ,
\end{equation}
we obtain the following functional equation for $S(\lambda)$ at $\lambda > 0$:
\begin{equation}\label{S-equation}
S(\epsilon\lambda) = S(\bar{\epsilon}\lambda).
\end{equation}
The function $S(\lambda)$ is also supposed to be an analytic
function at $\Re \lambda\geq 0$. Equation (\ref{S-equation})
requires $S(\lambda)$ to be an analytical function of $z=\lambda^4$:
\begin{equation}\label{S-series}
S(\lambda) = \sum_{n=0}^{\infty}s_n\lambda^{4n} \, .
\end{equation}
Therefore, being regular in the domain $\Re \,\lambda > 0$ (it is
even sufficient to require analyticity within a sector
$|\mathrm{arg}(\lambda)| \leq \pi/4$), the function $S(\lambda)$
must be regular on all the complex plane $\lambda$, i.e. it should
be an {\it entire function}. Now we apply the condition of
convergence of the integral over $\lambda$ in Eq.(\ref{canon}) at
large $\lambda$ to find the allowed asymptotic behavior of
$S(\lambda)$ at $|\lambda| \rightarrow\infty$. Substituting
Eq.(\ref{gen-C}) into Eq.(\ref{canon}) and using the asymptotics of
the Whittaker and $\Gamma$-functions we find that the integrand
behaves as $\lambda^{-3}S(\lambda)$ at $\lambda\rightarrow\infty$.
This means that $|S(\lambda)|$ should increase not faster than
$\lambda^{2}$. There is only one such entire function with the
structure of Eq.(\ref{S-series}): this is a constant
$S(\lambda)=s_{0}={\rm const}$. Thus we have proven  the uniqueness
of the solution Eq.(\ref{C0-solution}) up to a constant factor. This
factor has to be determined from the relation Eq.(\ref{Phi(0)}) and
the normalization condition for the phase distribution function
$\mathcal{P}(\phi)$. Now we may write down the solution for the
\emph{anomalous} (at the center of the band) generating function
$\Phi^{(an)}(u,\phi)$ in the final form :
\begin{eqnarray}\label{canon-final}
\hspace{-1.5cm} \Phi^{(an)}(u, \phi)
=\frac{u^{1/2}}{2\Gamma^4\left(\frac{1}{4}
\right)|\cos{\phi}\,\sin{\phi}|^{1/2}} &&\int_{0}^{\infty}d\lambda\,
\frac{\Gamma\left(\frac{1}{4}+\epsilon\lambda
\right)\,\Gamma\left(\frac{1}{4}+\bar{\epsilon}\lambda \right)}{\lambda^{3/2}} \nonumber \\
&& \left[W_{-\lambda\epsilon,\frac{1}{4}}\,\left(
\frac{\bar{\epsilon}\xi}{4\lambda}\right)W_{-\lambda\bar{\epsilon},\frac{1}{4}}\,\left(
\frac{\epsilon\eta}{4\lambda}\right)+ c.c. \, \right] \, ,
\end{eqnarray}
where $\xi = u\cos^2{\phi}$, $\eta= u\sin^2{\phi}$; $\epsilon =
\mathrm{e}^{i\pi/4}$, $\bar{\epsilon} = \mathrm{e}^{-i\pi/4}$.   In
the appendix we demonstrate explicitly that the obtained solution
Eq.(\ref{canon-final}) does obey the relation $\Phi^{(an)}(u=0,
\phi) = \mathcal{P}^{(an)}(\phi)$, where $\mathcal{P}^{(an)}(\phi)$
is given by Eq.(\ref{P0-solution}).
\begin{figure}[t]
\includegraphics[width=8cm, height=8
cm,angle=0]{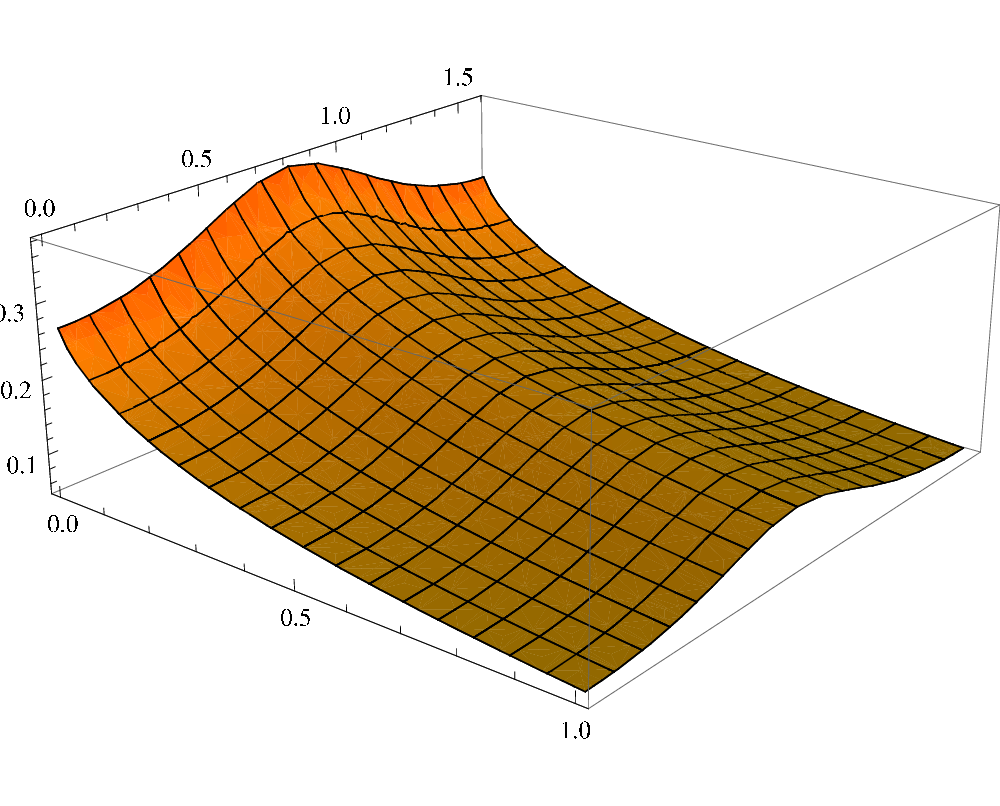} \caption{(color online) The function
$\Phi^{(an)}(u,\phi)$, in the range $u\in [0,1]$,
$\phi\in[0,\pi/2]$.}
\end{figure}
The 3D plot of the function $\Phi(u,\phi)$ is given in Fig.1. In the
next section we apply the solution Eq.(\ref{canon-final}) for
studying moments of the wave function distribution.

\section{Moments of the wave function distribution}\label{Physical applications}
The exact expression Eq.(\ref{canon-final}) for the anomalous (at
the center of the band) generating function is our main analytic
result. It determines statistics of wave function distribution at
the center-of-band anomaly. Although extensive physical applications
go beyond the framework of the present work, here we briefly discuss
the applicability of the one-parameter scaling description for the
anomalous statistics.

As has been mentioned in the introduction, the Lyapunov exponent
$\gamma(E)$, Eq.(\ref{Lyapunov}), sharply decreases in a narrow
vicinity of the band center $E=0$ \cite{Wegner, Derrida}. Being
dependent only on the phase distribution function, the Lyapunov
exponent can be easily calculated using  Eq.(\ref{P0-solution})  at
the center-of-band anomaly  $E=0$ and the trivial homogeneous phase
distribution $\mathcal{P}^{(norm)}(\phi) = 1/\pi$ close to but
outside the anomalous region. The ratio of real parts of the two corresponding
Lyapunov exponents $\gamma_{an}(E=0)$ and $\gamma_{norm}(E\approx
0)$ is given by \cite{Wegner, Derrida}:
\begin{equation}\label{Lyapunov-ratio}
\frac{\Re \, \gamma_{an}(E=0)}{\Re \, \gamma_{norm}(E \approx 0)}=
\int_{0}^{\pi}[1+\cos(4\phi)]\,\mathcal{P}^{(an)}(\phi)=\frac{8\,\Gamma^{2}\left(
\frac{3}{4}\right)}{\Gamma^{2}\left( \frac{1}{4}\right)}\approx
0.9139.
\end{equation}
According to Eq.(\ref{Lyapunov}), this can be interpreted as an
increasing localization length at the anomaly: $\ell_{0} \rightarrow
\ell_{an}^{{\rm ext}} = 1.094 \, \ell_{0}$. Note that the
localization length $\ell^{{\rm ext}}=1/(\Re \, \gamma)$ defined via
the Lyapunov exponent $\gamma$ characterizes the exponentially
decaying tails of the localized wave function and for this reason
will be referred to as the "extrinsic" localization length. In
contrast to that we consider the "intrinsic" localization length
$\ell^{{\rm int}}$ defined via the moments of the inverse
participation ratio
\begin{equation}
\label{int} \frac{I_{m}}{(m-1)!}=\frac{1}{(\ell^{{\rm int}})^{m-1}} \, ,
\end{equation}
whenever $\ell^{{\rm int}}$ is independent of $m$ in a sufficiently
wide interval of $m$.
This localization length characterizes the "body" of the localized
wave function.

Comparing Eqs.(\ref{mom-stand}) and (\ref{int}) one concludes that
away from $E=0$ anomaly the extrinsic and intrinsic localization
lengths coincide and are both equal to $\ell_{0}$ given by
Eq.(\ref{ell0}).  To study a relationship between them at the $E=0$
anomaly, we analyze the behavior of moments $I_m$ ($m>2$) of the
anomalous (at the center of band, $E=0$, $k=\pi/2$) wave function
distribution. With the definition Eq.(\ref{u-definition}), the
expression Eq.(\ref{IPR-m-integrals-z-phi-weak}) in the limit of a
long chain takes the form:
\begin{eqnarray}\label{IPR-m-integrals-u-phi}
\hspace{-2.5cm} I^{(an)}_{m}(E=0) =
\frac{4^{m}\pi\,\nu_{0}(0)}{(m-2)![\ell_{0}]^{m-1}\,\nu(0)}
\int^{\pi/2}_{0} d\phi \, \cos^{2m}({\phi}) \int^{\infty}_{0} d u \,
u^{m-2} [\Phi^{(an)}(u, \phi)]^2 \, ,
\end{eqnarray}
where the ratio $\nu_{0}(0)/\nu(0)$ is given by
Eq.(\ref{LDOS-z-phi-weak}):
\begin{equation}
\label{corr}
\frac{\nu(0)}{\nu_{0}(0)}=4\pi\int_{0}^{\pi/2}d\phi\,\cos^{2}(\phi)\,\,[\Phi^{(an)}(0,\phi)]^{2}.
\end{equation}

One might expect that the behavior of anomalous moments is similar
to Eq.(\ref{mom-stand}) but with the localization length $\ell_{0}$
replaced by some other length scale $\ell_{{\rm int}}$. This would
be the scenario of {\it one-parameter scaling} which appears to fail
at the band center.

A convenient way to present the results is to plot the {\it reduced
moments} $R_{m} \equiv I^{(an)}_{m}(E=0)/I^{(norm)}_{m}(E \approx
0)$ (where $I^{(norm)}_{m}=(m-1)!/\ell_{0}^{m-1}$,
Eq.(\ref{IPR-m-integrals-z-phi-weak}), are the moments
away from the anomaly):
\begin{equation}
\label{q-mom} \hspace{-2.0cm} R_{m} =
\frac{4^{m}\,\pi\,\nu_{0}(0)}{\Gamma(m)\Gamma(m-1)\,\nu(0)}\int_{0}^{\infty}du\int_{0}^{\pi}d\phi\,\cos^{2m}(\phi)\,u^{m-2}\,
[\Phi^{(an)}(u,\phi)]^2.
\end{equation}
Equation Eq.(\ref{q-mom}) with $\nu_{0}(0)/\nu(0)$ taken from
Eq.(\ref{corr}) and $\Phi^{(an)}(u,\phi)$ given by
Eq.(\ref{canon-final}) is parameter-free.

First of all we check that $R_{m\rightarrow 1}=1$ as normalization
of wave functions requires. Applying to Eq.(\ref{q-mom}) the
relation (with $\delta=m-1$):
\begin{equation}
\label{ident-delta}
\int_{0}^{\infty}dx\,x^{-1+\delta}\,f(x)=\delta^{-1}\,f(0)+O(1)\,\,\,\, ,
\,\, \delta=m-1\rightarrow 0 \, ,
\end{equation}
and using Eq.(\ref{corr}) one immediately obtains
$I^{(an)}_{m}=R_{m}=1$.\\
\begin{table}
\begin{tabular}{|c|c|c|c|c|c|c|c|c|c|}
               \hline
               1 & 2 & 3 & 4 & 5 & 6 & 7 & 8 & 9 & 10 \\
               1.0000 & 0.8347 & 0.6703 & 0.5321 & 0.4252 &0.3467  & 0.2908 &0.2519  & 0.2255 & 0.2083 \\
               \hline
             \end{tabular}\caption{Reduced moments $R_{m}$ ($m=1,2...10$) at the $E=0$ anomaly for an infinite
             chain in the limit of weak disorder.}
             \end{table}
The moments $I_{m}$ with $m>0$, $m\neq 1$ are essentially governed
by the $u$-dependence of the generating function
$\Phi^{(an)}(u,\phi)$. We evaluated numerically the reduced moments
$R_{m}$ up to $m=10$. The results are given in Fig.2 and Table 1.
\begin{figure}[t]
\includegraphics[width=6cm, height=8
cm,angle=-90]{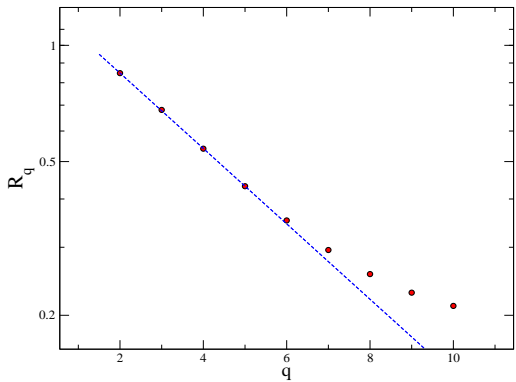} \caption{(color line)Reduced moments $R_{m}$
(red points) in the log-linear scale. The dashed line is the
exponential fit $R_{m}=(\ell_{0}/\ell^{{\rm int}}_{an})^{m-1}$ with
$\ell^{{\rm int}}_{an}/\ell_{0}=1.252$.}
\end{figure}

One can see that the behavior of reduced moments $R_m$ with
relatively small $m$ ($2\leq m<6$), $R_{m} \approx (\ell_{0}/\ell^{{\rm
int}}_{an})^{m-1}$, is, indeed, compatible with Eq.(\ref{int}). The
best exponential fit gives the same tendency of increasing the
localization length at the anomaly as in Eq.(\ref{Lyapunov-ratio}).
However, at the $E=0$ anomaly the extrinsic and intrinsic
localization lengths are no longer equal to each other, as it is the
case at energies away from the anomaly:
\begin{equation}
\label{ext-int} \ell^{{\rm int}}_{an}\approx
1.252\ell_{0},\;\;\;\;\;\ell^{{\rm ext}}_{an}\approx 1.094\ell_{0}.
\end{equation}
Even more interesting phenomenon takes place for large moments. At
$m>6$ one can see a significant enhancement of the moments compared
to their value extrapolated from the exponential dependence of
$R_{m}$ at small $m$. A possible physical meaning of these new
regime is discussed in a short publication \cite{KY-PRB}.

\section{Conclusions and open problems}
Eq.(\ref{canon-final}) is the main result of the paper. It gives an
exact and unique stationary (i.e. site-independent) solution
(in quadratures) to the
\emph{partial} differential equation Eq.(\ref{anomal-equation}) for
the generating function $\Phi(u, \phi)$ at the center of the energy
band ($k=\pi/2$, $E(k)=0$) of a weakly disordered chain. The
variables  $u$ and $\phi$ are associated with slowly varying
(squared) amplitude and phase of wave functions.  The generating
function we obtained can be used to compute all local statistics of
normalized eigenfunctions in the one-dimensional Anderson model in
the bulk of a long ($N\gg \ell$) chain. The solution of this problem
goes beyond the known problem of the Lyapunov exponent and related
quantities (e.g. density of states and conductance)
\cite{Wegner,Derrida,Titov,AL}, which
are completely determined by the distribution function of phase
$\mathcal{P}(\phi)$. As we have shown, $\mathcal{P}(\phi)$ is a
descender of the generating function $\Phi(u,\phi)$ and is related
with it in a simple way: $\mathcal{P}(\phi) = \Phi(u=0, \phi)$ (see
Eq.(\ref{Phi(0)})). Another important application of $\mathcal{P}(\phi)$
is that $\mathcal{P}(\phi=\theta/2)/2$ is the distribution function
$P_{\mathrm{ref}}(\theta)$ of a random phase $\theta$ of the coefficient of reflection
from a semi-infinite disordered chain.

The integrability of the partial differential equation
Eq.(\ref{anomal-equation-zero-mode}) for the generating function $\Phi(u,
\phi)$ which we discovered, is a remarkable evidence of a hidden
symmetry of the problem at $k=\pi/2$. Although  in the course of
derivation we mentioned  about the $\emph{sl}_{2}$ algebra of
operators Eq.(\ref{A-xy}), which are the building blocks for
Eq.(\ref{stat-xy}), and about even more extended algebra
Eq.(\ref{algebra-extended}), we did not exploit this algebraic
content explicitly. What would be highly useful is to find the
symmetry transformation which enables the discovered
separation of variables.
Moreover, the hidden symmetry which survives  violation of the
chiral (or sublattice) symmetry \cite{Dyson,Mirlin2008} by  on-site
disorder, could be important for other systems (like edge states in
Quantum Hall effect or in topological insulators) where the chiral
symmetry is broken by disorder. Speculating on its nature we may
surmise that this symmetry might be more naturally formulated in the
three dimensional space rather than in the two-dimensional space
$(\xi,\eta)$ and that it may have something to do with the symmetry
of the 3d harmonic oscillator. This conjecture is fed by an analogy
between our main result Eq.(\ref{canon-final}) and the expression for the
Green's function of the 3d harmonic oscillator problem \cite{BV}.
The analogy concerns the parameter ($\lambda$ in our problem and $k$
in Ref.\cite{BV}) entering both the argument and the first
index of the Whittaker functions in a mutually reciprocal way, as
well as the second index of the Whittaker functions being
$\frac{1}{4}$ in both cases (showing a connection with the
parabolic cylinder function). Establishing this symmetry would also
be useful for studying higher order anomalies at $k=\pi m/n$ with
$n>2$ (for the preliminary analysis of the ``devil's staircase"
of these higher-order anomalies, see Ref.\cite{K-Y-2008}).

We would like to note that the studied anomalies are inherent not
only to the problem of a disordered chain but might occur in other
physical situations where there is a periodic perturbation of random
amplitude. We mention an analogy between the 1d localization and the
classical system of kicked oscillator studied recently in
Ref.\cite{Izrail}. According to this analogy the energy-dependent
de-Broglie wavelength $\lambda_{E}$ of a particle on a chain is
encoded in the frequency of the oscillator while the lattice
constant $a$ determines the period of the $\delta$-function pulses
of the external force ("kicks"), their amplitude being proportional
to disorder. From this point of view, statistical anomalies arise
due to sequences of several kicks with amplitudes correlated in
time. Correlated amplitudes of kicks correspond to exclusive
configurations of the local disorder, hence the anomalies are weak
(for weak disorder) and exhibit themselves only in narrow windows
around selected energies. Remarkably, the variables $\xi$ and
$\eta$, which allowed us to factorize the equation for the
generating function of the Anderson model, play a role of the
co-ordinate and the momentum of the kicked oscillator.

Finally, we applied the exact solution for the generating function
to analyze the behavior of moments of the eigenfunction distribution
at the center-of-band anomaly. We have found that relatively small
moments behave similar to those outside the anomalous region but
with a renormalized localization length
$\ell_{0}\rightarrow\ell^{{\rm int}}_{an}$, while the larger moments
deviate significantly from this dependence. This fact together with the
appreciable enhancement of the "intrinsic" length $\ell^{{\rm int}}_{an}
\approx 1.252 \, \ell_{0}$ with respect to the
"extrinsic" length (inverse Lyapunov exponent)
$\ell^{{\rm ext}}_{an} \approx 1.094 \, \ell_{0}$, implies a significant
change
of the form of the "average" eigenfunction at the center-of-band
anomaly and simultaneously a failure of one-parameter description of
eigenfunction statistics.

\subsection*{Acknowledgments}
We appreciate stimulating discussions with A.Agrachev, B.L.Altshuler,
Y.V.Fyodorov, A.Kamenev, A.Ossipov, O.Yevtushenko, and a support from
RFBR grant 09-02-1235 (V.Y.). We are especially grateful to E.Cuevas
and D.N.Aristov for a help in numerical calculations.
A part of work was done during our stay at Kavli Institute for Theoretical Physics
at Santa Barbara and during visits of V.Y. to the Abdus Salam International
Center for Theoretical Physics. Support of these institutions is highly acknowledged.

\appendix\section{Normalization of $\Phi^{(an)}(u=0, \phi)$: explicit check of the relation $\Phi^{(an)}(u=0, \phi) =
\mathcal{P}^{(an)}(\phi)$ for Eqs. (\ref{canon-final}) and
(\ref{P0-solution}).} In order to take the limit $u \rightarrow 0$
in Eq.(\ref{canon-final}), we re-scale the integration variable
$\lambda \rightarrow u\lambda$ and use the identity $W_{0, \,
\nu}(z) = \sqrt{\frac{z}{\pi}}K_{\nu}(\frac{z}{2})$. Introducing a
new integration variable $x = 1/(8\lambda)$, we arrive at the
following expression for $\Phi^{(an)}(u=0, \, \phi)$:
\begin{eqnarray}\label{canon-final-u0}
\Phi^{(an)}(u=0, \,\phi) = \frac{4}{\pi\Gamma^2\left(\frac{1}{4}\right)}
\sqrt{|\sin{(2\phi)}|}\, \Re \,I
\, ,
\end{eqnarray}
where (see \cite{GR})
\begin{eqnarray}
I &\equiv& \int_{0}^{\infty}\sqrt{x}dx
K_{\frac{1}{4}}(\bar{\epsilon}x\cos^2\phi) K_{\frac{1}{4}}(\epsilon x \sin^2\phi) \label{Integral} \\
&=& \frac{\sqrt{2}\pi^2 i}{\Gamma^2\left(\frac{1}{4}\right)}\, \frac{\sin^{\frac{1}{2}}{\phi}}{|\cos{\phi}|^\frac{7}{2}}\,
F(1, \, \frac{3}{4}, \, \frac{3}{2}; \, 1 + \tan^4{\phi}) \label{Hypergeometry}\, .
\end{eqnarray}
The latter expression is rather complicated. To find its real part one has to
use nontrivial identities for the hypergeometric function. Obviously this brute
force approach does not exploit efficiently the symmetry of the problem.

It is more advantageous and instructive to exploit the symmetry and
perform the integration in Eq.(\ref{Integral}) in several elementary
steps. Using the representation (\cite{GR})
\begin{eqnarray}\label{Representaton}
K_{\frac{1}{4}}(z) = \left(\frac{z}{2}\right)^{1/4}
\frac{\Gamma\left(\frac{1}{2}\right)}
{\Gamma\left(\frac{3}{4}\right)}
\int^{\infty}_{0} dt \,\, \mathrm{e}^{-z \cosh{t}}\sqrt{\sinh{t}} \, ,
\end{eqnarray}
one can perform an elementary integration over $x$ in Eq.(\ref{Integral}) arriving at
\begin{eqnarray}\label{Integral-t1-t2}
\Re I = &&\frac{\Gamma^2\left(\frac{1}{4}\right)}{2\sqrt{2}\pi}|\sin{\phi}\cos{\phi}|^{1/2}
\int^{\infty}_{0} dt_1 \, dt_2 \sqrt{\sinh{t_1}\sinh{t_2}} \, \nonumber \\
&&\Re \, \frac{1}{[\,\bar{\epsilon}\cos^2{\phi}\,\cosh{t_1} + \epsilon \sin^2{\phi}\,\cosh{t_2}]^2}
\, .
\end{eqnarray}
Using the identity
\begin{eqnarray}\label{Identity}
\Re \, \frac{1}{[\,\bar{\epsilon} x  + \epsilon y]^2} = \frac{2xy}{[x^2 + y^2]^2}
\,
\end{eqnarray}
(for real $x$ and $y$) and introducing new variables:
\begin{eqnarray}\label{y-variables}
y_1 = \cos^2{\phi} \, \sinh{t_1} \,\,\,; \,\,\, y_2 = \sin^2{\phi} \, \sinh{t_2} \, ,
\end{eqnarray}
we obtain:
\begin{eqnarray}\label{Integral-y1-y2}
\Re I = \frac{\Gamma^2\left(\frac{1}{4}\right)}{\pi \, \sqrt{|\sin{(2\phi)}|}}
\int^{\infty}_{0} \frac{\sqrt{y_1y_2}\,dy_1 \, dy_2 }
{[\, \cos^4{\phi}\, + \sin^4{\phi}\, + y^2_1 + y^2_2]^2}
\, .
\end{eqnarray}
Re-scaling the variables $y_{1(2)} = y'_{1(2)}\sqrt{\cos^4{\phi} + \sin^4{\phi}}$ and introducing the
polar coordinates $\rho$ and $\alpha \in (0, \pi/2)$ as $y'_1 = \rho  \cos{\alpha}$ and $y'_2 =  \rho  \sin{\alpha}$, we get
\begin{eqnarray}\label{Integral-polar}
\Re I = \frac{\Gamma^2\left(\frac{1}{4}\right)}{\pi \, \sqrt{|\sin{(2\phi)}|}}
\frac{ I_{\alpha}I_{\rho}}{\sqrt{\cos^4{\phi}\, + \sin^4{\phi}\,}}
\, ,
\end{eqnarray}
where
\begin{eqnarray}\label{Simple integrals}
I_{\alpha} &=& \int^{\pi/2}_0 \sqrt{\sin{\alpha} \, \cos{\alpha}} = \frac{1}{2}B(\frac{3}{4}, \, \frac{3}{4}) =
\frac{2\pi^{3/2}}{\Gamma^2\left(\frac{1}{4}\right)} \, ; \\
I_{\rho} &=& \int^{\infty}_0 \frac{\rho^2 \, d\rho}{[1 + \rho^2]^2} = \frac{\pi}{4}
\, .
\end{eqnarray}
Collecting things together we arrive at the following final expression for $\Phi^{(an)}(u=0, \phi)$, Eq.(\ref{canon-final-u0}),:
\begin{eqnarray}\label{Phi-u0-answer}
\hspace{-2cm}
\Phi^{(an)}(u=0, \phi) = \frac{2\sqrt{\pi}}{\Gamma^2\left(\frac{1}{4}\right)}
\frac{1}{\sqrt{\cos^4{\phi}\, + \sin^4{\phi}\,}} = \frac{4\sqrt{\pi}}{\Gamma^2\left(\frac{1}{4}\right)}
\frac{1}{\sqrt{3 + \cos{(4\phi)}}}
\, .
\end{eqnarray}
This expression coincides with the anomalous probability
distribution of phase $\mathcal{P}^{(an)}(\phi)$
Eq.(\ref{P0-solution}) and thus proves the correct choice of the
numerical pre-factor in Eq.(\ref{canon-final}).


\section*{References}
\begin{thebibliography}{99}
\bibitem{Anderson} P.W.Anderson, Phys.Rev. \textbf{109}, 1492 (1958).
\bibitem{Mirlin2008} F.Evers and A.D.Mirlin, Rev.Mod.Phys \textbf{80}, 1355 (2008)
\bibitem{Mirlin2000} A.D.Mirlin, Phys.Rep.  \textbf{326}, 259 (2000).
 \bibitem{Modugno} G.Modugno, Rep.Prog.Phys., {\bf 73}, 102401 (2010).
\bibitem{Borland} R.E.Borland, Proc. R. Soc. A \textbf{274}, 529 (1963)
\bibitem{Halperin} B.I.Halperin, Phys.Rev. \textbf{139}, A104 (1965);
        Adv.Chem.Phys. \textbf{13}, 123 (1967).
\bibitem{Ber} V.L.Berezinskii, Zh.Exp.Teor.Fiz. {\bf 65}, 1251
(1973)[Sov.Phys.JETP {\bf 38}, 620 (1974)].
\bibitem{AR} A.A.Abrikosov and I.A.Ryzhkin, Adv.Phys. {\bf 27}, 147
(1978).
\bibitem{Mel} V.I.Melnikov, JETP Lett. {\bf 32}, 225 (1980).
\bibitem{Kolok} I.V.Kolokolov, Zh.Exp.Teor.Fiz. {\bf 103}, 2196
(1993)[JETP {\bf 76}, 1099 (1993)].
\bibitem{Pendry-rev} J.B.Pendry, Adv.Phys. \textbf{43}, 461 (1994).
\bibitem{1dRev} J.Frohlich, F.Martinelli, E.Scoppola and T.Spencer,
Comm.Math.Phys. {\bf 101}, 21 (1985).
\bibitem{Pastur} I.M.Lifshitz, S.A.Gredeskul and L.A.Pastur,
{\it Introduction to the theory of disordered systems} (Wiley, New
York, 1988).
\bibitem{Wegner}M.Kappus and F.Wegner, Z.Phys. B {\bf 45}, 15 (1981).
\bibitem{Derrida} B.Derrida and E.Gardner, J.Phys. (Paris) {\bf 45}, 1283
(1984).
\bibitem{Izrail} L.Tessieri and F.M.Izrailev, Phys.Rev.E {\bf 62}, 3090 (2000).
\bibitem{Titov} H.Schomerus and M.Titov, Phys.Rev.B {\bf 67}, 100201(R) (2003).
\bibitem{AL} L.I.Deych, M.V.Erementchouk, A.A.Lisyansky, and B.L.Altshuller,
 Phys.Rev.Lett. {\bf 91}, 096601 (2003).
\bibitem{Efet-book} K.B.Efetov, {\it Supersymmetry in chaos and disorder} (Cambridge University Press, Cambridge, England, 1977).
\bibitem{Mirlin-Fyodorov-1991} A.D.Mirlin and Y.V.Fyodorov, Nucl. Phys. B \textbf{366}, 507 (1991).
\bibitem{A-C-Anderson-Thouless-1973} R.Abou-Chacra, P.W. Anderson, and D.J. Thouless, J. Phys. C
\textbf{6}, 1734 (1973).
\bibitem{OsK} A.Ossipov and V.E.Kravtsov, Phys.Rev.B {\bf 73}, 033105 (2006).
\bibitem{rem1} All the positive moments of this distribution are
finite, in particular $\langle |\psi|^{2} \rangle=\frac{1}{N}$.
\bibitem{GR} I.S.Gradshtein and I.M.Ryzhik {\it Table of integrals series and products}
(Academic Press, New York, 1996).
\bibitem{Barnes&Luck-1990} C.Barnes and J.M.Luck, J.Phys. A \textbf{23}, 1717 (1990).
\bibitem{BV} V.L.Bakhrakh and S.I.Vetchinkin, Theor.Math.Phys. {\bf 6}, 283 (1971) [Sov. Phys:
Teor. Mat. Fiz.{\bf 6}, 392 (1971)].
\bibitem{K-Y-2008} V.E.Kravtsov and V.I.Yudson,
  AIP Conference Proceedings, {\bf 1134}, 31 (2009). ArXiv:0806.2118;
\bibitem{Dyson} F.J.Dyson, Phys.Rev. {\bf 92}, 1331 (1958).
\bibitem{KY-PRB} V.E.Kravtsov and V.I.Yudson, Phys.Rev. B \textbf{82}, 195120 (2010).
\bibitem{Landau-QM} L.D.Landau and E.M.Lifshitz, Quantum Mechanics, Elsevier Science Ltd. 1977.
\end{thebibliography}
\end{document}